\documentclass[prb,twocolumn,letterpaper,superscriptaddress]{revtex4-2}
\usepackage[space]{grffile}
\usepackage{bm,amsmath,color}
\usepackage{multirow,tabularx}
\usepackage{longtable}
\usepackage{stix,pifont}
\usepackage{mathrsfs}
\usepackage{xcolor}
\usepackage{float}  
\usepackage{changes}
\usepackage{enumitem}
\graphicspath{{./figs/png/}}

\def\version{2.9}

\def\Vbckgnd{\ensuremath \overline{\mathscr{V}}_2}
\def\UppRi{\ensuremath \text{\rotatebox[origin=c]{0}{$\circleurquadblack$}}} \def\UppLe{\ensuremath \text{\rotatebox[origin=c]{90}{$\circleurquadblack$}}} \def\BotLe{\ensuremath \text{\rotatebox[origin=c]{180}{$\circleurquadblack$}}} \def\BotRi{\ensuremath \text{\rotatebox[origin=c]{270}{$\circleurquadblack$}}}

\begin{document}

\title{Nonlocal magnon transconductance in extended magnetic insulating films.\\II: two-fluid behavior.} 

\author{R. Kohno}
\affiliation{Université Grenoble Alpes, CEA, CNRS, Grenoble INP, Spintec, 38054 Grenoble, France}

\author{K. An}
\affiliation{Université Grenoble Alpes, CEA, CNRS, Grenoble INP, Spintec, 38054 Grenoble, France}

\author{E. Clot}
\affiliation{Université Grenoble Alpes, CEA, CNRS, Grenoble INP, Spintec, 38054 Grenoble, France}

\author{V. V. Naletov} 
\affiliation{Université Grenoble Alpes, CEA, CNRS, Grenoble INP, Spintec, 38054 Grenoble, France}

\author{N. Thiery}
\affiliation{Université Grenoble Alpes, CEA, CNRS, Grenoble INP, Spintec, 38054 Grenoble, France}

\author{L. Vila}
\affiliation{Université Grenoble Alpes, CEA, CNRS, Grenoble INP, Spintec, 38054 Grenoble, France}

\author{R. Schlitz}
\affiliation{Department of Materials, ETH Zürich, 8093 Zürich, Switzerland}

\author{N. Beaulieu} 
\affiliation{LabSTICC, CNRS, Universit\'e de Bretagne Occidentale,
  29238 Brest, France}

\author{J. Ben Youssef} 
\affiliation{LabSTICC, CNRS, Universit\'e de Bretagne Occidentale,
  29238 Brest, France}
  
\author{A. Anane} 
\affiliation{Unit\'e Mixte de Physique CNRS, Thales, 
    Univ. Paris-Sud, Universit\'e Paris Saclay, 91767 Palaiseau, France}

\author{V. Cros} 
\affiliation{Unit\'e Mixte de Physique CNRS, Thales, 
    Univ. Paris-Sud, Universit\'e Paris Saclay, 91767 Palaiseau, France}

\author{H. Merbouche} 
\affiliation{Unit\'e Mixte de Physique CNRS, Thales, 
    Univ. Paris-Sud, Universit\'e Paris Saclay, 91767 Palaiseau, France}

\author{T. Hauet} 
\affiliation{Université de Lorraine, CNRS Institut Jean Lamour, 
    54000 Nancy, France}

\author{V. E. Demidov}
\affiliation{Department of Physics, University of Muenster, 48149 Muenster, Germany}

\author{S. O. Demokritov} 
\affiliation{Department of Physics, University of Muenster, 48149 Muenster, Germany}

\author{G. de Loubens} 
\affiliation{SPEC, CEA-Saclay, CNRS, Universit\'e Paris-Saclay,
  91191 Gif-sur-Yvette, France}

\author{O. Klein}
\email[Corresponding author:]{ oklein@cea.fr}
\affiliation{Université Grenoble Alpes, CEA, CNRS, Grenoble INP, Spintec, 38054 Grenoble, France}

\date{\today}

\begin{abstract}
This review presents a comprehensive study of the spatial dispersion of propagating magnons electrically emitted in extended yttrium-iron garnet (YIG) films by the spin transfer effects across a YIG$\vert$Pt interface. Our goal is to provide a generic framework to describe the magnon transconductance inside magnetic films. We experimentally elucidate the relevant spectral contributions by studying the lateral decay of the magnon signal. While most of the injected magnons do not reach the collector, the propagating magnons can be split into two-fluids: \textit{i)} a large fraction of high-energy magnons carrying energy of about $k_B T_0$, where $T_0$ is the lattice temperature, with a characteristic decay length in the sub-micrometer range, and \textit{ii)} a small fraction of low-energy magnons, which are particles carrying energy of about $\hbar \omega_K$, where $\omega_K/(2 \pi)$ is the Kittel frequency, with a characteristic decay length in the micrometer range. Taking advantage of their different physical properties, the low-energy magnons can become the dominant fluid \textit{i)} at large spin transfer rates for the bias causing the emission of magnons, \textit{ii)} at large distance from the emitter, \textit{iii)} at small film thickness, or \textit{iv)} for reduced band mismatch between the YIG below the emitter and the bulk due to variation of the magnon concentration. This broader picture complements part I \cite{kohno_SD}, which focuses solely on the nonlinear transport properties of low-energy magnons.

\end{abstract}
\maketitle

\begin{figure}
    \includegraphics[width=0.49\textwidth]{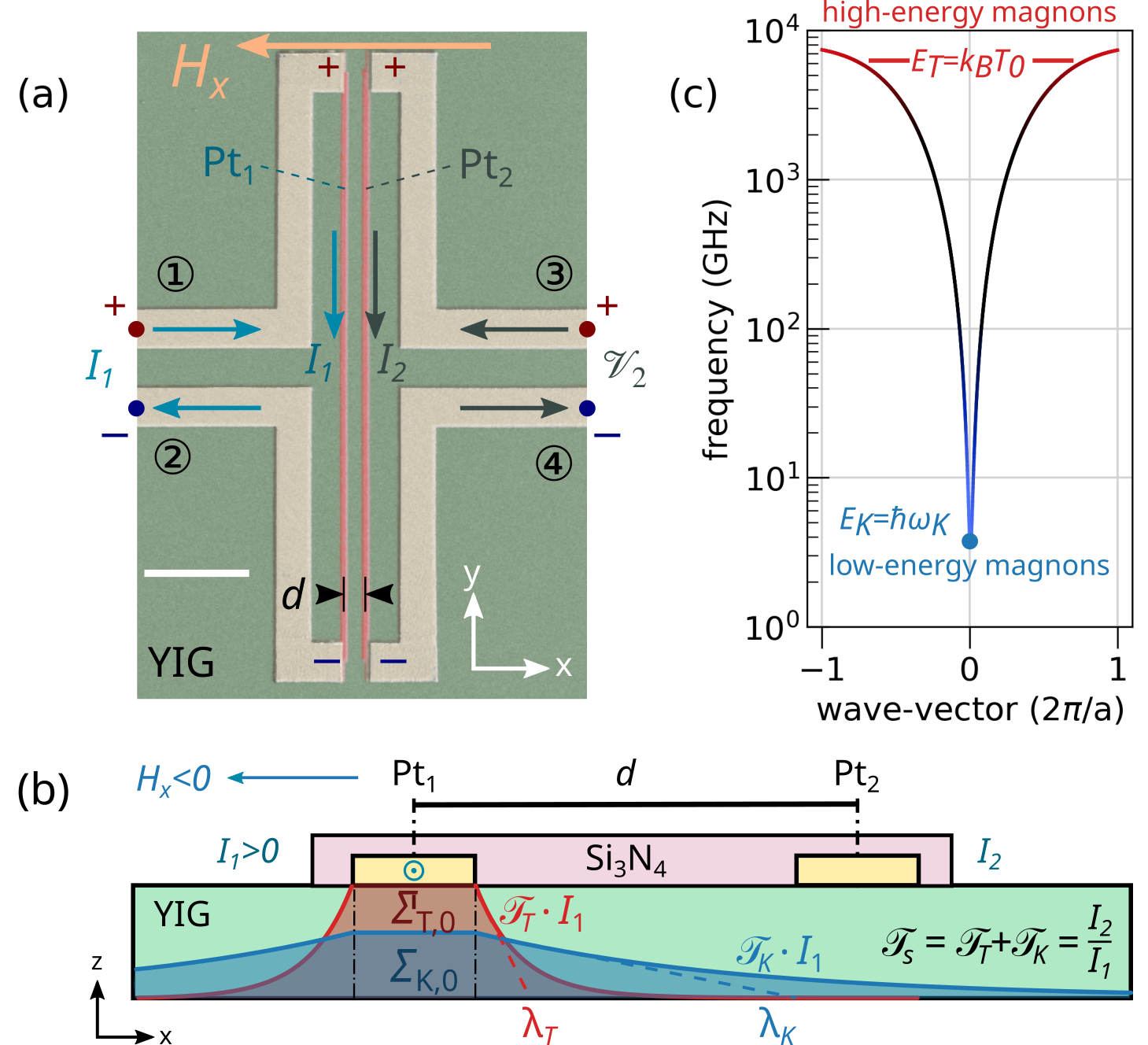}
    \caption{Lateral geometry used for measuring the magnon transconductance in extended magnetic insulating films. (a) Scanning electron microscope image of a 4-terminal circuit (scale bar is 5~$\mu$m), whose 4 poles are connected to two parallel wires, Pt$_1$ and Pt$_2$ (shown in pink), deposited on top of a continuous YIG thin film. A continuous electric current, $I_1$, injected in Pt$_1$ (emitter) produces an electric modulation of the magnon population by the spin transfer effect (STE). This modulation is consequently detected laterally by the spin pumping voltage $-R_2 I_2$ through a second electrode Pt$_2$ (collector) placed at a distance $d$ from the emitter. We define the magnon transmission ratio $\mathscr{T}_s=I_2/I_1$ and the transconductance $\mathscr{T}_s/R_1$.  Panel (b) is a sectional view showing the spatial decay of propagating magnons. (c) Schematic representation of the spin-wave dispersion over the Brillouin zone. We consider the spin transport properties as originating from two independent fluids located at either end of the dispersion curve.  Each of the two-fluids has a different characteristic decay length, $\lambda_T$ and $\lambda_K$ respectively, as shown in (b).}
    \label{fig:intro}
\end{figure}
 
\section{Introduction} \label{sec:intro}


Nonlocal devices, such as the geometry shown in Fig.~\ref{fig:intro}, consisting of two lateral circuits deposited on an extended magnetic insulating film have recently attracted much attention as novel electronic devices exploiting the spin degree of freedom\cite{Cornelissen2015,Goennenwein2015,thiery2018,Lebrun2018,Brataas2020spin}. As emphasized in part I, one of their original features is to behave as a spin diode at large currents\cite{kohno_SD}. These devices rely on the spin transfer effect (STE) to electrically modulate the magnon population in a magnetic thin film. The process alters the amplitude of thermally activated spin fluctuations by transferring quanta of $\gamma \hbar$ between an adjacent metallic electrode and the magnetic thin film via a stimulated emission process. In unconfined geometries, a wide energy range of eigenmodes is available to carry the external flow of angular momentum, spanning a frequency window from GHz to THz, as schematically shown in Fig.~\ref{fig:intro}(c), which shows the lower branch of the spin wave dispersion over the Brillouin zone \cite{barker2016thermal,princep2017full,nambu2020observation}. At high-energy the curve flattens out at about 30~meV, which corresponds to the thermal energy, $E_T \approx k_B T_0$, at ambient temperature, while at low-energy it shows a gap, $E_g \approx \hbar \omega_K \approx 30$~$\mu$eV, around the Kittel frequency $\omega_K/(2 \pi)$ \cite{gurevich2020magnetization}. Between these two extremes, the spectral identification of the relevant eigenmodes involved in nonlocal spin transport has remained mostly elusive.

In this review, we propose a simple analytical framework to account for the magnon transconductance in extended magnetic insulating films. We find that the observed behavior can be well approximated by a two-fluid model, which simplifies the spectral view as emanating from two independent types of magnons placed at either end of the magnon manifold. On the one hand, we have magnons at thermal energies, to be referred to as \emph{high-energy} magnons\cite{thiery2018}, whose distribution follows the temperature of the lattice. On the other hand, we have magnons at the bottom of the band near the Kittel frequency, to be referred to as \emph{low-energy} magnons, whose electrical modulation at high power is the focus of part I\cite{kohno_SD}. The response of these two magnon populations to external stimuli is very different. The high-energy thermal magnons, being particles of high wavevector, are mostly insensitive to any changes in the external conditions of the sample such as shape, anisotropy and magnetic field, being instead defined by the spin-wave exchange stiffness and the large k-value of the magnon\cite{etesami2015spectral,Adachi2013}. In contrast, low-energy magnons, sensitive to magnetostatic interaction, depend sensitively on the extrinsic conditions of the sample. It turns out that nonlocal devices provide a unique means to study each of these two-fluids independently by comparing the differences in transport behavior as a function of the separation, $d$, between the two circuits, thus benefiting from the spatial filtering associated with the fact that each of these two components decays very differently as a function of distance, as schematically shown in Fig.~\ref{fig:intro}(b).

The paper is organized as follows. After this introduction, in the second section we review the main features that support the two-fluid separation. In the third section, we describe the analytical framework of a two-fluid model and, in particular, the expected signature in the transport measurement. This part builds on the knowledge gained in part I\cite{kohno_SD} about the nonlinear behavior of the low-energy magnon. To facilitate quick reading of either manuscript, we point out that a summary of the highlights is provided after each introduction and, in both papers, the figures are organized into a self-explanatory storyboard, summarized by a short sentence at the beginning of each caption. In the fourth section we will show the experimental evidence that supports such a picture and finally in the fifth section we will conclude our work by emphasizing the important results and opening to future perspectives.

\section{Key findings} \label{sec:key}

The purpose of this review is to present the experimental evidence supporting the separation of the magnon transconductance into two components. This is achieved by measuring the transmission coefficient $\mathscr{T}_s \equiv I_2/I_1$ of magnons emitted and collected via the spin Hall effect between two parallel Pt wires, Pt$_1$ and Pt$_2$, respectively. It is shown that a two-fluid model, where $\mathscr{T}_s=\mathscr{T}_T+\mathscr{T}_K $ is the independent sum of a high-energy and a low-energy magnon contribution, provides a simplified common framework that captures all the observed behavior in nonlocal devices with different inter-electrode separation, different current bias, different applied magnetic field, different film thickness or magnetic composition, and different substrate temperature. 

Making a quantitative analysis of the transmission ratio, we find that most of the injected spins remain localized under the emitter or propagate in the wrong direction (the estimated fraction is about 2/3), making these materials intrinsically \emph{poor} magnon conductors. The remaining propagating magnons fall into two distinct categories: First, a large fraction carried by high-energy magnons, which follow a diffusive transport behavior with a characteristic decay length, $\lambda_T$, in the submicron range\cite{an2021short,jamison2019long}; and second, a small fraction carried by low-energy magnons, which are responsible for the asymmetric transport behavior \cite{kohno_SD}, and which follow a ballistic transport with a characteristic decay length, $\lambda_K$, in the micrometer range. The different decay behaviors are directly observable experimentally in the change of the nonlinear spin transport behavior with separation, $d$.

We also carefully study the collapse of the magnon transmission ratio with increasing temperature of the emitter, $T_1$, as it approaches the Curie temperature, $T_c$. Here, the number of spin-polarized sites under the electrode becomes of the same order as the spin flux coming from the external Pt electrode. The transition to this regime of magnetization reduction leads to a sharp decrease in the magnon transmission ratio. We report signs of interaction between the low-energy and high-energy parts of the liquid in this highly diffusive regime\cite{bender2016thermally,flebus2016local,yu2017thermal}. In addition, the collapse seems to actually occur well before reaching $T_c$, suggesting that the total number of magnons is significantly underestimated compared to that inferred from the single temperature value of the lattice below the emitter. Alternatively, this discrepancy could indicate a rotation of the equilibrium magnetization under the emitter\cite{ulrichs2020chaotic,avci2017current}. Since the discrepancy actually becomes more pronounced as the magnetic film gets thinner, this suggests that the culprit is the amount of low-energy magnons.

\section{Analytical framework} \label{sec:analytical}

\begin{figure}
    \includegraphics[width=0.49\textwidth]{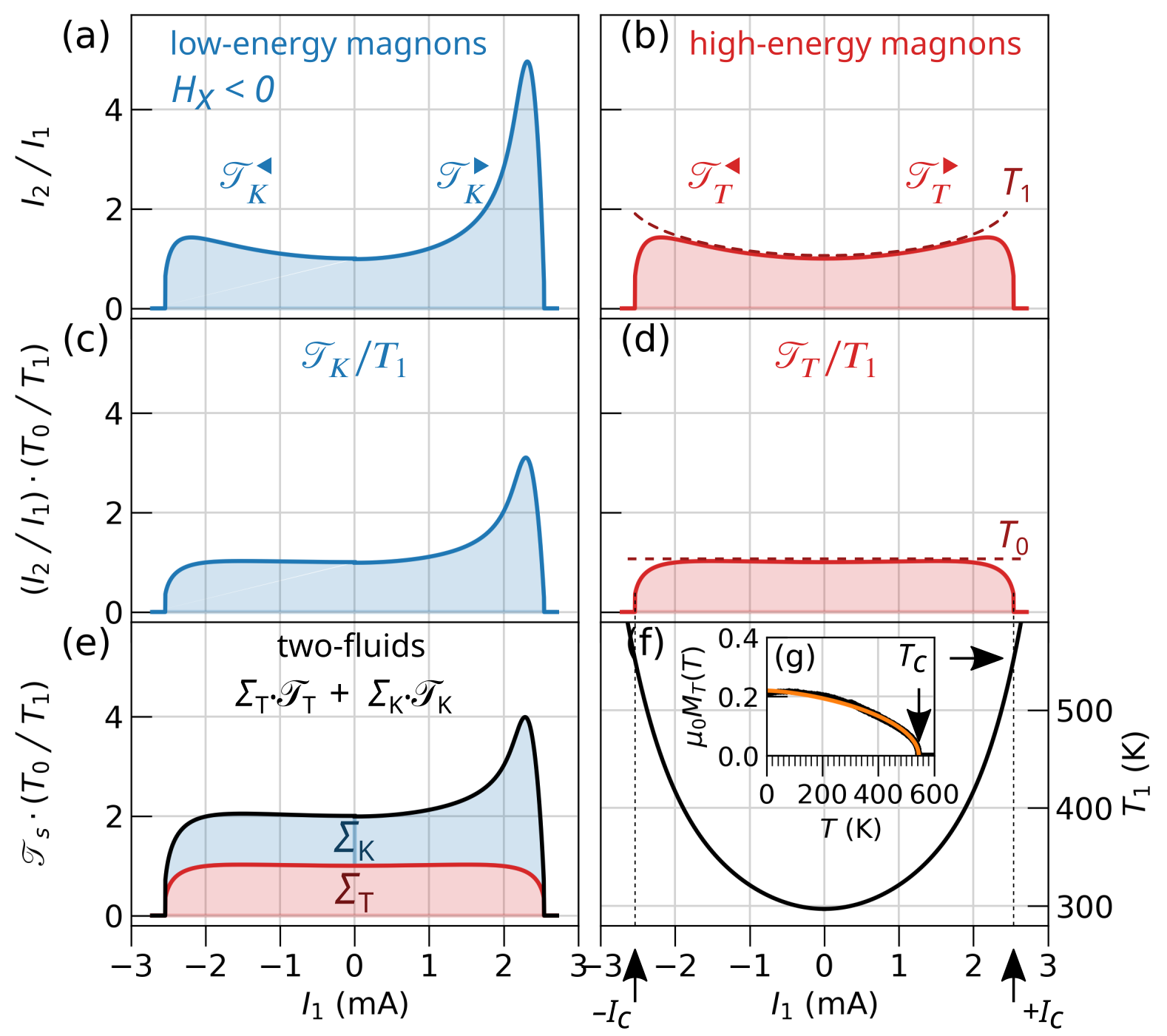}
    \caption{Current bias characteristic of the magnon transconductance depending on the spectral nature of propagating magnons. Panels (a) and (b) compare the predicted electrical variation of $\mathscr{T}_s$ for low-energy magnons (see Eq.~(\ref{eq:dnk}), left panel) and for high-energy magnons (see Eq.~(\ref{eq:dnt}), right panel), respectively, when $H_x <0$.  Panel (f) shows the associated variation of $T_1= T_0 + \kappa R I_1^2$, the lattice temperature below the emitter. The current span exceeds $I_\text{c}$, the current bias, which raises $T_1$ to $T_c$, the Curie temperature. Panels (c) and (d) show the behavior when $\mathscr{T}_s$ is renormalized by $T_1$. Panel (e) shows the two-fluid fitting function: the independent sum of the low-energy and high-energy magnon contributions with their respective weights $\Sigma_T$ and $\Sigma_K$.  
    The inset (g) shows the temperature dependence of the magnetization $M_T$ as measured by vibrating sample magnetometry (cf. Fig.~\ref{fig:carac}), and the solid line is a fit with the analytical expression $M_T \approx M_0 \sqrt{1 - (T/T_c)^{3/2}}$, with $\mu_0 M_0=0.21$~T and $T_c=550$~K. }
    \label{fig:dn}
\end{figure}

\subsection{Low-energy magnons}

We recall the finding in part I\cite{kohno_SD}, that the transconductance by low-energy magnons in open geometries can be described by the analytical expression:
\begin{equation}
     \mathscr{T}_K  \propto  \dfrac{M_1}{M_2} \cdot \dfrac {k_B T_1}{\hbar \omega_K}  \cdot  \dfrac  {e \omega_K} {\mathcal{I}_\text{th}} \cdot  \dfrac 1 {1 - \left ( I_1 / {\mathcal{I}_\text{th}} \right )^2}\hspace{0.2cm}, \label{eq:dnk}
\end{equation} 
where $e$ is the electron charge, while $M_1$ and $M_2$ are the magnetization values under the emitter and collector, respectively. The threshold current, $\mathcal{I}_\text{th}$, is the solution of a transcendental equation obtained by combining Eqs. (4), (6) and (7) in Ref.~\cite{kohno_SD}. In our model, its nonlinear behavior is determined solely by two parameters $\mathcal{I}_{\text{th},0}$ and $n_\text{sat}$, which are related to the nominal value of the transmission coefficient at low current and the saturation threshold expressed in normalized units of nonlinear effects, respectively. All information about these feedback effects can be found in Ref.~\cite{kohno_SD}.

As emphasized in detail in part I, one of the pitfalls of nonlocal devices is that the emitter electrode cannot be made immune to Joule heating due to poor thermalization in the 2D geometry. This leads to a significant increase of the temperature under the emitter with current $I_1$, which we model by 
\begin{equation}
\left . T_1 \right |_{I_1^2} \; = \; T_0 + \kappa \,R \, I_1^2. \label{eq:temp}
\end{equation}
In our notation, $T_0$ is the substrate temperature at no current and $\kappa$ is the temperature coefficient of resistance for Pt. It is the coefficient that determines the temperature rise per deposited joule power (see Fig.~S1 in Appendix). We additionally define $I_\text{c}$ the current required to reach the Curie temperature, $T_c=T_0 + \kappa R I_\text{c}^2$ [see Fig.~\ref{fig:dn}(f)].  This variation has profound consequences both on the level of thermal fluctuations of the low-energy magnons and on the number of high-energy magnons. In particular, the variation of $T_1$ with $I_1$ expressed by Eq.~(\ref{eq:temp}) enters into the variation of $\mathscr{T}_K$ with $I_1$ expressed by Eq.~(\ref{eq:dnk}).  The resulting variation of the magnon population as a function of $I_1$ is shown in Fig.~\ref{fig:dn}(a). To account for the variation of $T_1$ produced by Joule heating, which expresses the influence of a varying background of thermal fluctuations on the STE, we plot $\mathscr{T}_K/T_1$ in Fig.~\ref{fig:dn}(c). This renormalization is equivalent to looking at the nonlinear behavior from the perspective of a thermalized background. The resulting shape of the curve as a function of $I_1$ is greatly simplified. In the reverse bias, marked by the symbol $\smallblacktriangleleft$ representing the magnon absorption regime, the normalized transconductance is constant up to $I_\text{c}$. In contrast, in the forward bias, denoted by the symbol $\smallblacktriangleright$, which represents the magnon emission regime, a peak appears. This asymmetric peak is called the spin diode effect in part I\cite{kohno_SD}. The advantage of the $T_1$ normalization of the magnon transmission ratio is that it makes the peak a characteristic feature of the spin diode effect. 

\subsection{High-energy magnons}

We now assume that the number of high-energy magnons is approximately equal to the total number of magnons, which is the difference $M_1 - M_0$, where $M_0$ is the spontaneous magnetization at $T=0$~K and $M_1$ is the spontaneous magnetization at $T=T_1$, the temperature of the emitter \footnote{This approximation of neglecting the contribution of low-energy magnons to the total number of magnons is consistent with making the saturated magnetization, a quantity that counts the total number of magnons, a constant of motion when studying the high-power regime of magnetostatic modes.}. We thus analytically express the contribution of high-energy magnons to the magnon transconductance by the equation:
\begin{equation}
    \mathscr{T}_T \propto \dfrac{M_1}{M_2} \cdot \dfrac{M_0 - M_1}{M_0}\hspace{0.2cm}, \label{eq:dnt}
\end{equation}
where the prefactor $M_1$ represents the amount of magnetic polarization available under the emitter. We note that the analytical form expressed by Eq.~(\ref{eq:dnt}) has been previously proposed to describe spin transmission in paramagnetic materials\cite{Oyanagi2019}. As shown in the inset Fig.~\ref{fig:dn}(g), we find that the temperature dependence of $M_1$ is well described by the analytical $M_1 \approx M_0 \sqrt{1 - (T_1/T_c)^{3/2}}$. The resulting number of thermally excited magnons contributing to the nonlocal transport is shown in Fig.~\ref{fig:dn}(b). Repeating the same analysis developed in Fig.~\ref{fig:dn}(c), a more revealing behavior is obtained by renormalizing $\mathscr{T}_T$ with $T_1$ and the result is shown in Fig.~\ref{fig:dn}(d). In this case, the current dependence of $\mathscr{T}_T/T_1$ on $I_1$ is a constant function up to $I_\text{c}$.

\subsection{Two-Fluid Model}

\begin{figure}
    \includegraphics[width=0.49\textwidth]{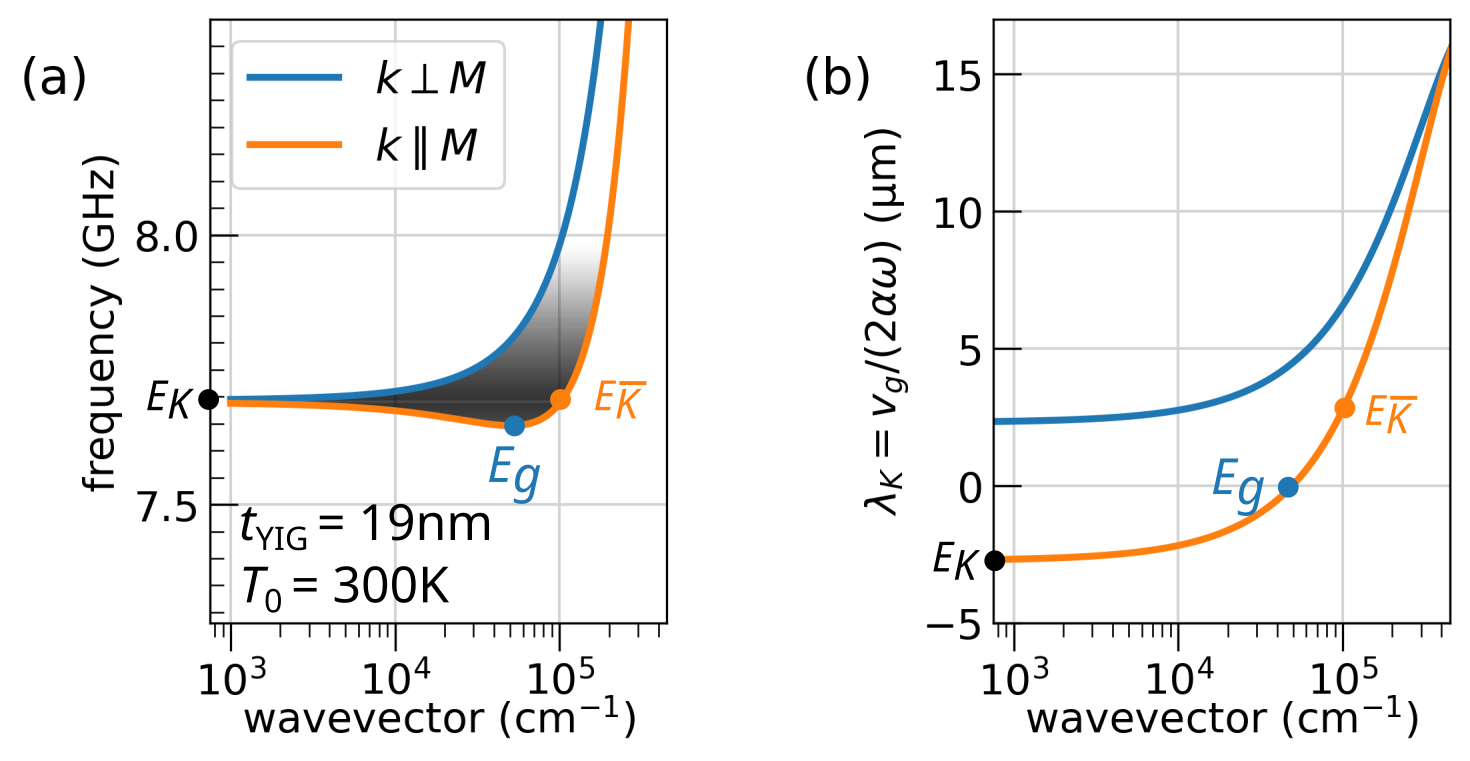}
    \caption{Dispersion characteristic of low-energy magnons. (a) Dispersion curves at the bottom of the magnon manifold of a 19~nm thick YIG film for two values of $\theta_k=0^\circ$ ($k \parallel M$) and $90^\circ$ ($k \perp M$), the angle between the wavevector and the applied magnetic field. We mark with dots the Kittel mode ($E_K$, black dot), the lowest energy mode ($E_g$, blue dot), and the mode degenerate to the Kittel mode with the highest wavevector ($E_{\overline{K}}$, orange dot). The curve is computed for YIG$_A$ thin films. (b) Characteristic decay length calculated from the dispersion curve, assuming that the magnons follow the phenomenological LLG equation with $\alpha_\text{LLG} = 4 \cdot 10^{-4}$.}
    \label{fig:vg}
\end{figure}

An advantage specific to nonlocal transport measurements is that the propagation distance, $d$, provides a powerful means to spectrally distinguish different types of magnons, each of which has its characteristic decay length $\lambda_{k}$ along the $x$-axis \cite{an2021short,gomez2020differences}. In the following we will examine the expectation for the different extrema of the dispersion curve.

For the high-energy magnons, the spin wave spectrum can simply be approximated as $\omega_k = \omega_M \lambda_\text{ex}^2 k^2 $, where $\omega_M = \gamma \mu_0 M_s = 2\pi \times 4.48$~GHz and $\lambda_\text{ex} \approx 15$~nm is the exchange length\cite{Cherepanov1993}. High-energy magnons at room temperature ($T_0 = 300$~K) have the frequency $\omega_T = k_B T_0 /\hbar = 2\pi \times 6.25$~THz, which corresponds to a wavevector $k_T = 2.5$~nm$^{-1}$. It is seriously questionable whether the estimate for $\lambda_T$ from the phenomenological Landau-Lifshitz-Gilbert (LLG) model is applicable to such short-wavelength magnons.  Practically \textit{i)} the Gilbert damping is expected to be increased in the THz range \cite{Cherepanov1993}. \textit{ii)} the group velocity is reduced towards the edge of the Brillouin zone \cite{plant1977spinwave,barker2016thermal}, and \textit{iii)} the LLG model does not consider the reduction of the characteristic propagation distance due to diffusion processes. Furthermore, YIG is a ferrimagnet, higher (antiferromagnetic) spin wave branches contribute significantly to the magnon transport\cite{barker2016thermal,princep2017full,nambu2020observation}. We believe that the most reliable estimates have been obtained experimentally by studying the spatial decay of the spin Seebeck signal\cite{uchida2008observation,an2021short} and have found $\lambda_T \approx 0.3$~$\mu$m.

In contrast to its high-energy counterpart, the LLG framework should provide a good basis for calculating the propagation distance of long-wavelength dipolar spin waves. This interaction gives an anisotropic character to the group velocity of these spin waves. In Fig~\ref{fig:vg}(a) we plot the dispersion curve of a magnon propagating either along the $x$-axis (orange line) or along the $y$-axis (blue line). In the following, we will focus our attention on the branch $\theta_k = 0^\circ$ (orange line), which corresponds to the magnon propagating in the normal direction of the Pt wires. As emphasized in part I\cite{kohno_SD}, there are 3 remarkable positions on the curve, each marked by a colored dot on Fig.~\ref{fig:vg}. The energy minimum, $E_g$ (blue dot), does not contribute to the transport because its group velocity is zero. The longest wavelength spin waves correspond to the Kittel mode, $E_K$ (black dot). The damping rate, taking into account the ellipticity of the spin waves, is given by $\Gamma_K = \alpha_\text{LLG} (\omega_H + \omega_M/2)$, where $\omega_H = \gamma H_0$\cite{Loubens2005}. The velocity is equal to $v_K = \partial_k \omega= \omega_H \omega_M t_\text{YIG} / (4 \omega_K)$, where $\omega_K$ is the Kittel frequency and $t_\text{YIG}$ is the YIG thickness. The resulting decay length of the spin transport carried by $k\rightarrow0$ magnons is $\lambda_K=v_K/(2\Gamma_K) \approx 2.5$~$\mu$m for $t_\text{YIG}=19$~nm. As pointed out in part I\cite{kohno_SD}, the mode that seems to be most relevant for long-range magnon transport in nonlocal devices is probably $E_{\overline{K}}$, the degenerate mode with the Kittel frequency and the shortest wavelength. This mode is marked by an orange dot in Fig.~\ref{fig:vg}. For our $t_\text{YIG}=19$~nm film, it turns out that its group velocity is of the same order as that of the Kittel mode, giving a similar decay distance. We will show later that this estimate is quite close to the experimental value. We note, however, that the value of the decay distance at $E_{\overline{K}}$ increases with increasing film thickness to become independent of $t_\text{YIG}$ for thicknesses above 200~nm. The saturation value is $\lambda_K \approx 20$~$\mu$m, assuming $\alpha_\text{LLG} = 4 \cdot 10^{-4}$.

Since $\lambda_K \approx 10 \times \lambda_T$, changing $d$ allows tuning from spin transport governed by high-energy magnons to spin transport governed by low-energy magnons. One should also add that the current intensity, $I_1$, also provides a means to tune the ratio between the two-fluid as discussed in Ref.~\cite{kohno_SD}. 

Learning from the above considerations, we can now put all the contributions together to propose an analytical fit of the data with the two-fluid function:
\begin{equation}
   {\mathscr{T}_s} = \Sigma_{T,0} \; \exp^{-d/\lambda_T} \dfrac {\mathscr{T}_T}{\mathscr{T}_{T,I_1 \rightarrow 0}} + \Sigma_{K,0} \; \exp^{-d/\lambda_K}   \; \dfrac {\mathscr{T}_K}{\mathscr{T}_{K,I_1 \rightarrow 0}} \hspace{0.2cm}, \label{eq:2fluid}
\end{equation}
combining two independent magnon contributions: one at thermal energy and the second at magnetostatic energy. We assume here that both magnon fluids follow an exponential decay. To ease the notation, we shall refer below at underlined quantity, \textit{e.g.} $\underline{\mathscr{T}}_T \equiv {\mathscr{T}_T}/{\mathscr{T}_{T,I_1 \rightarrow 0}} $, as the normalized quantity by the low current value. We define $\left .\Sigma_{K} \right |_d = \Sigma_{K,0} \; \exp^{-d/\lambda_K}$ and $\left .\Sigma_{T} \right |_d = \Sigma_{T,0} \; \exp^{-d/\lambda_T}$, where the index $0$ represents the extrapolated value at the emitter position ($d=0$): see Fig.~\ref{fig:intro}(b). Thus the parameter $\left . \Sigma_{K}/(\Sigma_{K}+\Sigma_{T}) \right |_d$ represents the variation with distance of the proportion of low-energy propagating magnons over the total number of propagating magnons. An exemplary fit for $d=0$ and identical high-energy and low-energy contributions is shown in Fig.~\ref{fig:dn}(e).

It should be emphasized that the model proposed by Eq.~(\ref{eq:2fluid}), which assigns a fixed decay rate to each magnon category, is certainly too simplistic. For example, one should keep in mind that if $M_1 \rightarrow 0$ due to Joule heating, this could have profound consequences on $\lambda_T$ by changing the stiffness of the exchange constant. This has already been discussed in the context of spin propagation in paramagnetic materials\cite{Oyanagi2019}. We will return to this issue below in the context of our discussion of the discrepancy in the values of $T_c$ extracted from the transport data.

\section{Experiments} \label{sec:experiment}

In this section we present the experimental evidence supporting the two-fluid picture shown above. We focus on the evolution of spin transport with current, distance, applied magnetic field, substrate temperature and effective magnetization, $M_\text{eff}$. This will allow us to test the validity of our model. 

\begin{figure}
    \includegraphics[width=0.49\textwidth]{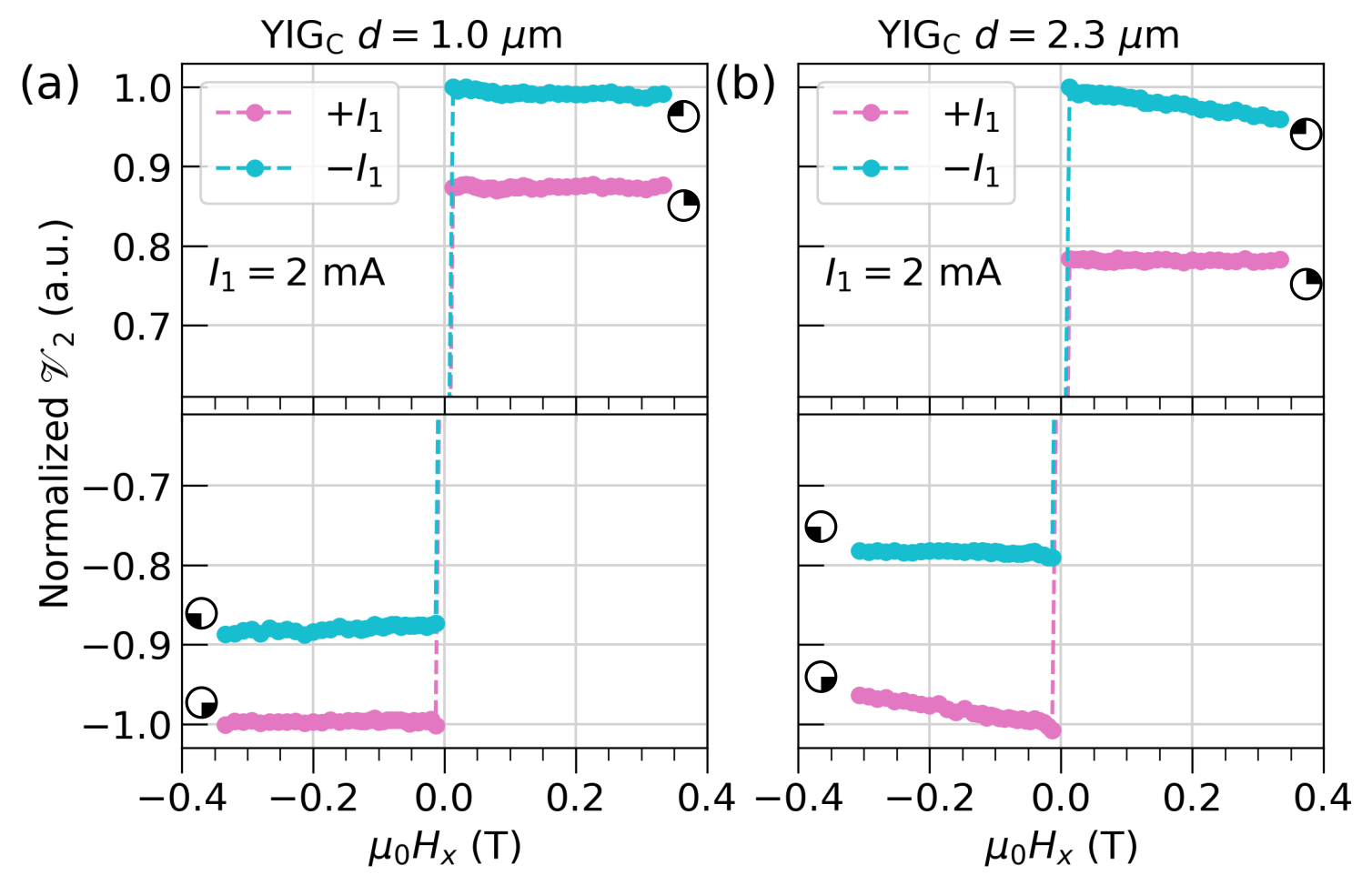}
    \caption{Dependence of the collected voltage on an external magnetic field. Comparison of the nonlocal voltage $\mathscr{V}_2 = (V_{2,\perp} - V_{2,\parallel})$ between (a) a short-range device ($d=1.0$~$\mu$m) and (b) a long-range device ($d=2.3$~$\mu$m). The panels show the zoom at the maximum and minimum of the normalized values. We interpret the detection of a finite susceptibility, $\partial_{H_x} \mathscr{V}_2 < 0$, as an indication of a magnon transmission ratio by low-energy magnons. In contrast, a constant behavior, $\partial_{H_x} \mathscr{V}_2 \approx 0$, is indicative of a magnon transmission ratio by high-energy magnons. Finite susceptibility is uniquely observed in the long-range regime when $I_1 \cdot H_x <0$, \textit{i)} when the number of low-energy magnons is increased by injecting a current in the forward direction, and \textit{ii)} when the contribution of the rapidly decaying high-energy magnons becomes a minority.  The data are collected on the YIG$_C$ thin film driven by a large current amplitude of $\pm I_1=2.0$ mA. The normalization value of $\mathscr{V}_2$ are respectively $10.04$~$\mu$V and $8.66$~$\mu$V in panel (a) and (b). 
    }
    \label{fig:hdep}
\end{figure}

\subsection{Magnetic susceptibility of the magnon transmission ratio} \label{sec:2fluids}

We begin this section by first presenting some key experimental evidence supporting the two-fluid picture. A schematic of the 4-terminal device is shown in Fig.~\ref{fig:intro}(a). It circulates pure spin currents between two parallel electrodes subject to the spin Hall effect\cite{sinova2015spin}: in our case two Pt strips $L_\text{Pt}=30$~$\mu$m long, $w_\text{Pt}=0.3$~$\mu$m wide and $t_\text{Pt}=7$~nm thick.  The experiment is performed here at room temperature, $T_0=300$~K, on a 56~nm thick (YIG$_C$) garnet thin film whose physical properties are summarized in Table~1 of Ref.~\cite{kohno_SD}. While injecting an electric current $I_1$ into Pt$_1$, we measure a voltage $V_2$ across Pt$_2$, whose resistance is $R_2$. To subtract all non-magnetic contributions, we define the spin signal $\mathscr{V}_2 = (V_{2,\perp} - V_{2,\parallel})$ as the voltage difference between the normal and parallel configuration of the magnetic field with respect to the direction of the electric current. In practice, the measurement is obtained simply by recording the change in voltage as an in-plane external magnetic field, $H_0$, is rotated along the $x$ and $y$ directions, respectively [the Cartesian frame is defined in Fig.~\ref{fig:intro}(a)]. Fig.~\ref{fig:hdep} shows the variation of $\mathscr{V}_2$ as a function of $H_x$ for a large amplitude of $|I_1|=$2.0~mA, which corresponds to a current density of $1 \cdot 10^{12}~$A/m$^2$. To reduce the influence of Joule heating and also thermal activation of the electrical carriers in YIG\cite{Thiery2018a,schlitz2021nonlocal}, we use a pulse method with a 10\% duty cycle throughout this study to measure the nonlocal voltage\cite{thiery2018}. 
In the measurements, the current is injected into the device only during 10~ms pulses with a 10\% duty cycle. 
In Fig.~\ref{fig:hdep} we compare the magnetic field sensitivity of the (normalized) spin transport at two values of the center-to-center distance $d$ between emitter and collector for positive and negative polarities of the current. In total, this leads to 4 possible configurations for the pair ($I_1$,$H_x$), each labeled by the symbols \rotatebox[origin=c]{0}{$\circleurquadblack$}, \rotatebox[origin=c]{90}{$\circleurquadblack$}, \rotatebox[origin=c]{180}{$\circleurquadblack$}, \rotatebox[origin=c]{270}{$\circleurquadblack$} to match the notation of Fig. ~\ref{fig:raw}. There, vertical displacement of the marker dissociates scans of opposite $H_x$-polarity, while horizontal displacement of the marker dissociates scans of opposite $I_1$-polarity. Looking at Fig.~\ref{fig:hdep}, we recover the expected inversion symmetry 
while enhancement of the spin current is clearly visible when $I_1 \cdot H_x <0$. 
The signal seems to depend on the magnetic field only for larger distances and $I_1 \cdot H_x <0$ (forward bias).  Considering that the two Pt wires are both $w=0.3$~$\mu$m wide, this corresponds to an edge to edge separation $s=d-w$. In one case the distance is $s \approx (2 \lambda_T)$, in the other case $s \approx 4 \cdot (2 \lambda_T)$, where $2 \lambda_T \approx 0.6 \mu$m is the estimated amplitude decay length of the magnons at thermal energy. It will be shown below that under the emitter the number of high-energy magnons far exceeds the number of low-energy magnons. Assuming an exponential decay of the high-energy magnons, one expects in (a) an attenuation of their contribution by 50\%, while in (b) it is reduced by almost 99\%. We thus arrive at a situation where at $d=0.5$~$\mu$m the magnon transport is dominated by the behavior of high-energy magnons, while at $d=2.3$~$\mu$m the magnon transport is dominated by the behavior of low-energy magnons (see below). In Fig.~\ref{fig:hdep} we assign the finite susceptibility $\partial_{H_x} \mathscr{V}_2 < 0$ as an indication of magnon transmission through low-energy magnons. Since the energy of these magnons as well as the threshold of damping compensation depend sensitively on the magnetic field\cite{Slavin09,hamadeh2012autonomous}, the low-energy magnons are significantly affected by the amplitude of the magnetic field, $H_x$ \cite{thiery2018,guckelhorn2021magnon,Wimmer2019}. Such a field dependence is explained in Eq.~(5) of Ref.~\cite{kohno_SD}. What is observed here is that near the peak bias, $I_\text{pk}\approx 2.2$~mA (see definition in part I), the device becomes particularly sensitive to a shift of $\mathcal{I}_\text{th}$. In our case, the external magnetic field shifts $\mathcal{I}_\text{th}$ by shifting the Kittel frequency, $\omega_K= \gamma \mu_0 \sqrt{H_0(H_0+M_s)}$.  In contrast, the constant behavior, $\partial_{H_x} \mathscr{V}_2 \approx 0$, is indicative of a magnon transmission ratio by high-energy magnons: because of their short wavelength, their energy is of the order of the exchange energy, and thus independent of the magnetic field strength\cite{uchida2010insulator}. Since these 2 plots are measured with exactly the same current bias, and the only parameter changed is $d$, it shows that filtering between high and low-energy magnons can be achieved by simply changing the separation between emitter and collector. It also directly suggests a double exponential decay, as will be discussed later in Fig.~\ref{fig:gap}. 

\subsection{Spectral signature in nonlocal measurement.}

\begin{figure}
    \includegraphics[width=0.49\textwidth]{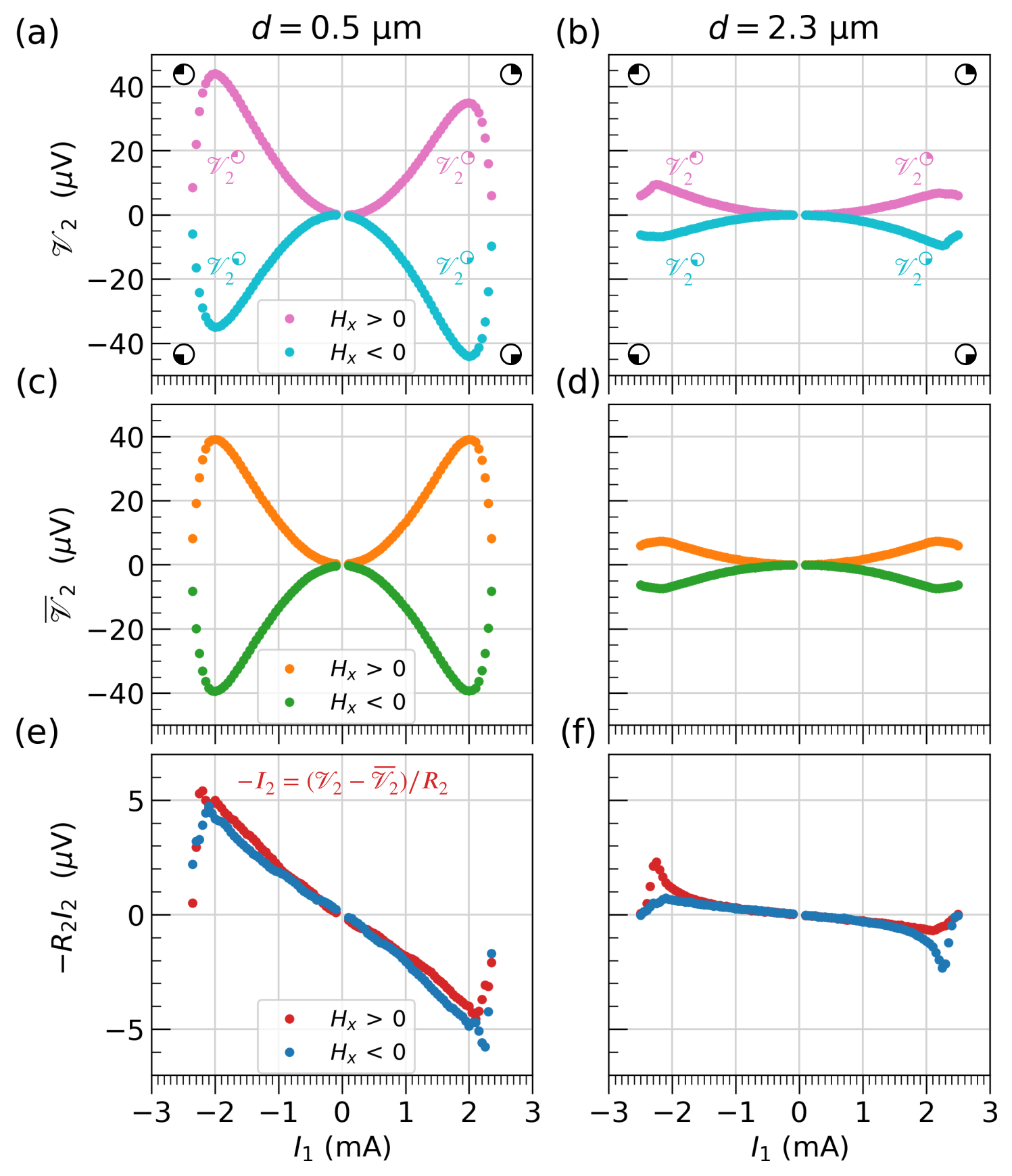} 
    \caption{Measurement of the collected electrical current, $I_2$, as a function of the emitter current, $I_1$. We compare the transport characteristics between two nonlocal devices: one with a short emitter-collector distance in the submicron range ($d=0.5$~$\mu$m, left column) and the other with a long distance of a few microns ($d=2.3$~$\mu$m, right column). The first row (a) and (b) shows $\mathscr{V}_2$ at $T_0=300$~K as a function of $I_1$, the injected current, for both positive and negative polarity of $H_x$, the applied magnetic field. In our symbol notation, the marker position indicates the quadrant in the plot pattern. The raw signal $\mathscr{V}_2 = -R_2 I_2 + \overline{\mathscr{V}_2}$ is decomposed into an electric signal, $I_2$, and a thermal background signal, $\overline{\mathscr{V}_2}$, as shown in the third row (e,f) and the second row (c,d), respectively. The background, $\overline{\mathscr{V}_2}$, represents the background magnon currents along the thermal gradients. The measurements are performed on YIG$_\text{C}$ thin films. The data are taken at $H_0=0.2$~T.}
    \label{fig:raw}
\end{figure}

Fig.~\ref{fig:raw} compares the variation of $\mathscr{V}_2$ as a function of emitter current $I_1$ for two different emitter-collector separations. The maximum current injected into the device is about 2.5~mA, corresponding to a current density of $1.2 \cdot 10^{12}~$A/m$^2$. The polarity bias for the pair ($I_1$,$H_x$) is represented by the symbols \rotatebox[origin=c]{0}{$\circleurquadblack$}, \rotatebox[origin=c]{90}{$\circleurquadblack$}, \rotatebox[origin=c]{180}{$\circleurquadblack$}, \rotatebox[origin=c]{270}{$\circleurquadblack$}, in replication of the 4-curve pattern. We recover in Fig.~\ref{fig:raw}(a,b) the expected inversion symmetry with $\mathscr{V}_2^\text{\rotatebox[origin=c]{0}{$\circleurquadblack$}} \approx -\mathscr{V}_2^\text{\rotatebox[origin=c]{180}{$\circleurquadblack$}}$ and $\mathscr{V}_2^\text{\rotatebox[origin=c]{90}{$\circleurquadblack$}} \approx -\mathscr{V}_2^\text{\rotatebox[origin=c]{270}{$\circleurquadblack$}}$, while the enhancement of the spin current is visible when $I_1 \cdot H_x <0$, representing the forward regime. As explained in part I \cite{kohno_SD}, the raw signal $\mathscr{V}_2 = \overline{\mathscr{V}_2} -R_2 I_2$ can be decomposed into \textit{i)} $\left . \overline{\mathscr{V}_2} \right |_{I_1^2}$ a thermal signal produced by the Spin Seebeck Effect (SSE), which is always odd/even with $H_x$ or $I_1$ and shown in panels (c,d), and \textit{ii)} $-R_2 \left . I_2 \right |_{I_1}$, an electrical signal produced by the spin transfer effect (STE), which is in the linear regime even/odd with the polarity of $H_x$ or $I_1$, respectively, and shown in panels (e,f) \footnote{The minus sign in front of the electrical contribution accounts for the fact that the spin-charge conversion is an electromotive force and thus the current flows in the opposite direction to the voltage drop.}. This decomposition is obtained by assuming that in reverse bias $\Vbckgnd^\UppRi = -\mathscr{V}_2^\BotLe + R_2 \left . \mathscr{T}_s \right |_{I_1 \rightarrow 0} \; \underline{\mathscr{T}}_T \cdot I_1 $ and $\Vbckgnd^\BotLe = -\mathscr{V}_2^\UppRi + R_2 \left . \mathscr{T}_s \right |_{I_1 \rightarrow 0} \; \underline{\mathscr{T}}_T \cdot I_1 $, which evaluates the number of absorbed magnons as a linear deviation from the number of thermally excited low-energy magnons, assuming C$^2$ continuity of the magnon transmission ratio across the origin. We recall that in our notation $\underline{\mathscr{T}}_T \equiv \mathscr{T}_T/\left . \mathscr{T}_T \right |_{I_1 \rightarrow 0} $.  We then construct $\Vbckgnd^\UppLe = \Vbckgnd^\UppRi$ and $\Vbckgnd^\BotRi = \Vbckgnd^\BotLe$ by enforcing that the signal generated by Joule heating is exactly even in $I_1$. We observe that in the short range ($d=0. 5$~$\mu$m), we get $\Vbckgnd^\circletophalfblack \approx (\mathscr{V}_2^\UppLe+\mathscr{V}_2^\UppRi )/2$ and $I_2^\circletophalfblack = \text{ sign} (I_1) (\mathscr{V}_2^\UppLe- \mathscr{V}_2^\UppRi )/(2 R_2)$, which is the expected signature for a symmetric magnon signal. This equality is not satisfied in the long range ($d=2.3$~$\mu$m) for $\Vbckgnd$ due to the asymmetry of the signal between forward and reverse bias as explained in part I. The consistency of this data manipulation is confirmed below in Fig. ~\ref{fig:sigma}(a) and (b) by showing a small asymmetric enhancement of $I_2$ at high $I_1$ by low-energy magnons at short distances and a pronounced enhancement at long distances as discussed in Ref.~\cite{kohno_SD}. The fact that a more pronounced enhancement is observed at large distances is further evidence for the spatial filtering of high-energy magnons.

It is worth noting that one can reach a situation where $-R_2 I_2=0$ without necessarily having $\overline{\mathscr{V}_2}$ vanish as well, as shown in Fig.~\ref{fig:raw}(e) and (f) at $I_1 =2.5$~mA. This is explained by the formation of lateral temperature gradients\cite{Shan2017}. In other words, the observation of $M_T=0$ is a local problem, mostly affecting the region below the emitter. It does not imply that $M=0$ throughout the thin film.

\begin{figure}
    \includegraphics[width=0.49\textwidth]{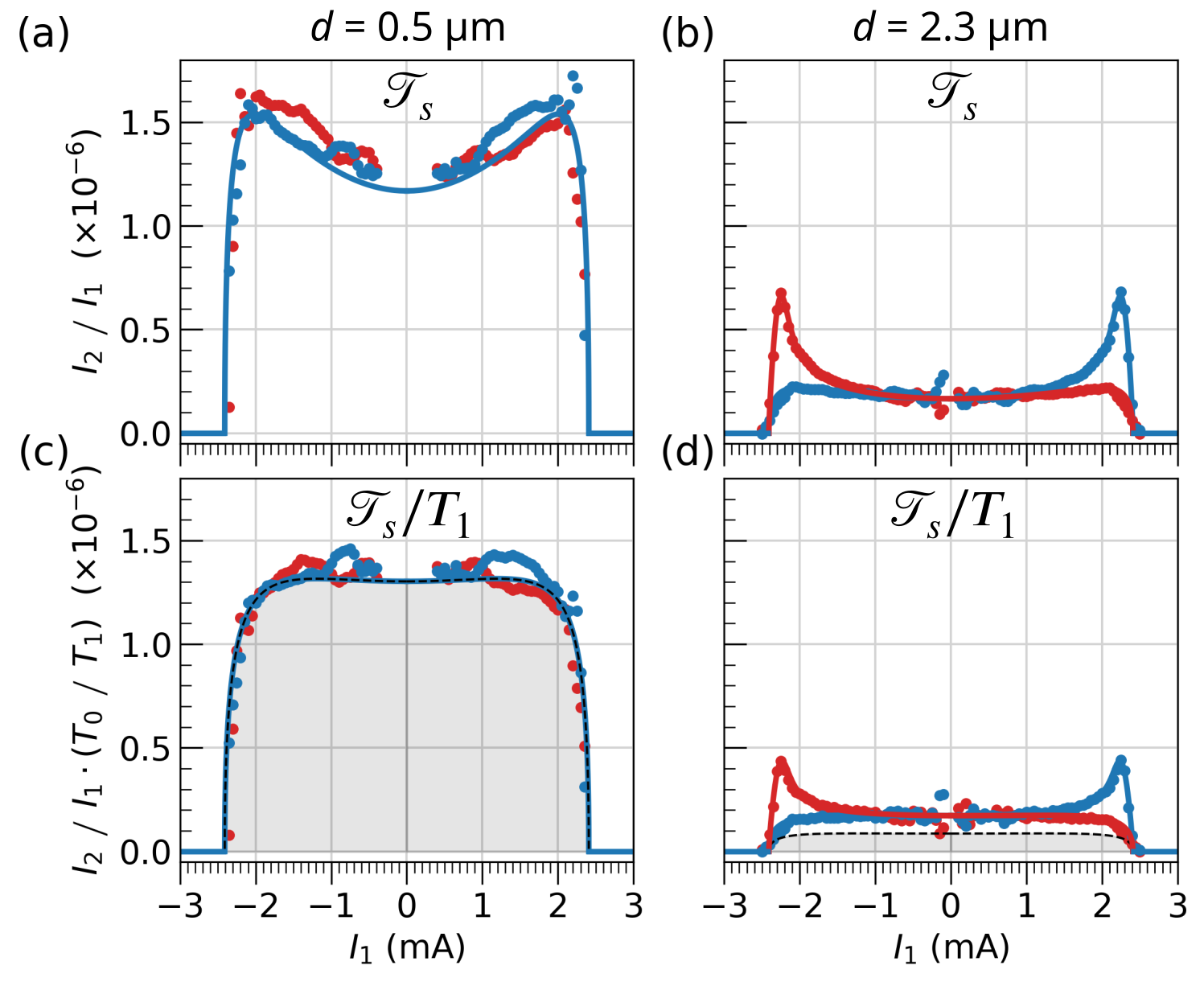} 
    \caption{Dependence of the magnon transmission ratio on the separation between the electrodes. Starting from the extraction of $I_2$ in Fig.~\ref{fig:raw}, the first row compares the variation of the ratio $\mathscr{T}_s= I_2/I_1$ between short-range (left column) and long-range (right column) devices. In the short range, the behavior shows a symmetrical signal of the magnon transmission ratio with respect to the current polarity $I_1$, while in the long range, the behavior is asymmetrical. We interpret the difference to be due to two different types of magnons: dominantly high-energy magnons in the short range and dominantly low-energy magnons in the long range. To eliminate nonlinear distortions caused by Joule heating, $\mathscr{T}_s$ is renormalized by $\left . T_1 \right |_{I_1^2}$, the emitter temperature variation produced by Joule heating (see text). The solid lines are fitted with Eq.~(\ref{eq:2fluid}), where the shaded region shows the background contribution from high-energy magnons $\Sigma_T \underline{\mathscr{T}}_T $, where $\left . \Sigma_T/(\Sigma_T+\Sigma_K) \right |_{d}$ represent their relative weight at this distance. In (c) this ratio is about 0.95, while in (d) it drops to about 0.5.}
    \label{fig:sigma}
\end{figure}

\begin{figure}
    \includegraphics[width=0.49\textwidth]{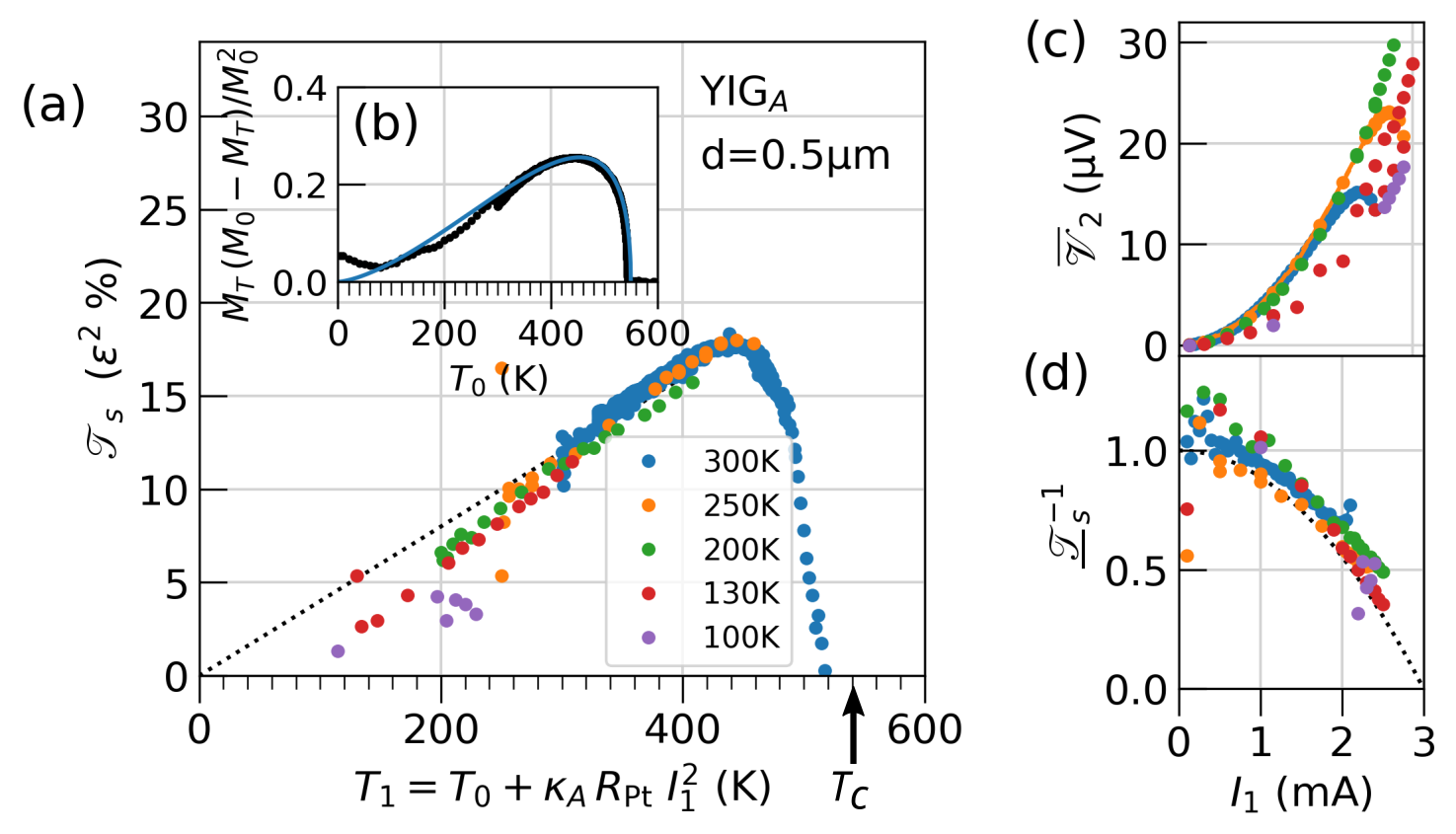}
    \caption{Dependence of the magnon transmission ratio on the substrate temperature, $T_0$. Short-range measurement ($d=0.5$~$\mu$m) of nonlocal spin transport in YIG$_\text{A}$. (a) Variation of $\mathscr{T}_s$ as the emitter current $I_1$ is varied in the range $[-2.5, 2.5]$~mA at different values of the substrate temperature $T_0$. The data are plotted as a function of $T_1=T_0+\kappa_A R_\text{Pt} I_1^2$, the emitter temperature. The resulting temperature dependence of $\mathscr{T}_s$ observed in (a) corresponds to the variation of $M_T (M_0-M_T)$ shown in inset (b), where $M_T$ is the temperature dependence of the saturation magnetization.  The dots are the experimental points, while the blue solid line is the expected behavior assuming $M_T \approx M_0 \sqrt{1 - (T/T_c)^{3/2}}$. This transformation is supported by the observation in (c) and (d) that both the SSE signal $\overline{\mathscr{V}_2}$ and the normalized inverse transmission ratio $\underline{\mathscr{T}}_s$ vs. $I_1$ scale on the same parabolic behavior (dashed line), suggesting that the relevant bias parameter is $T_1$.  }
    \label{fig:lsv3}
\end{figure}

As a next step, we will show how to distinguish the contributions of high-energy and low-energy magnons using the analytical model in Fig.~\ref{fig:sigma}. Starting from Fig.~\ref{fig:raw}(e,f), we will remove the influence of the spurious contribution on the electrical spin transport signal. First, we normalize the signal by the emitter current to obtain the magnon transmission ratio coefficient $\mathscr{T}_s= I_2/I_1$ as shown in Fig.~\ref{fig:sigma}(a,b). For small separation, we observe that $\mathscr{T}_s$ shows a quadratic behavior that is symmetric in current and consequently we associate it with the device temperature. In contrast, the device with large separation shows an asymmetric enhancement due to the spin diode effect~\cite{kohno_SD}. The influence of the increase of the emitter temperature $T_1$ due to the Joule heating of $I_1$ can be removed by normalizing with $T_1/T_0$. This normalization removes the symmetric enhancement of the magnon transmission ratio as reported in previous studies\cite{Zhang2012,Goennenwein2015,Cornelissen2016,schlitz2021nonlocal}, where the justification will be discussed later in Fig.~\ref{fig:lsv3} \footnote{Theoretical\cite{Zhang2012} and experimental studies\cite{Goennenwein2015} reported the power law of $T^{3/2}$ at low temperature $T < 300$~K. However, the exponent decreases to $T^{1/2}$ as the temperature increases. Finally, in our temperature range, the exponent $1$ fits the data well, as shown in Fig.~\ref{fig:lsv3}\cite{schlitz2021nonlocal}. }. The obtained traces are shown in Fig.~\ref{fig:sigma}(c,d) and can be compared with the theoretical expectation given by Eq.~(\ref{eq:2fluid}), which is graphically summarized in Fig.~\ref{fig:dn}(e). The solid lines are fit curves with our model representing the sum of the contribution from low-energy magnons and the background contributions from high-energy magnons, with the parameters of the fit given in Table~\ref{tab:fit}. The dashed line and the gray shaded area represent the latter $\Sigma_T \Delta n_T$. From the fits we can obtain the ratio $\Sigma_T/(\Sigma_T+\Sigma_K)$ for the two magnon fluids, where the contribution of high-energy magnons decreases from 95\% at 0.5$\mu$m to 50\% at 2.3$\mu$m, in accordance with the spatial filtering proposed above.

To illustrate Eq.~(\ref{eq:dnt}) experimentally, we repeated the measurement for different values of the substrate temperature $T_0$ at small separation. Fig.~\ref{fig:lsv3}(a) shows the experimental result for five different values of $T_0$ when $I_1$ varies on the same $[-2.5,2.5]$~mA span. Note that the data are plotted as a function of $T_1=T_0+\kappa_A R_\text{Pt} I_1^2$, the emitter temperature.  The rationale for this transformation of the abscissa is apparent in Fig.~\ref{fig:lsv3}(b) and (c), which show that the nonlinear current dependence of both the SSE and STE signals originates from the enhancement of $T_1$. In particular, Fig.~\ref{fig:lsv3}(c) shows the rise of the SSE signal $\overline{\mathscr{V}_2}$ as a function of $I_1$ for different values of $T_0$. We find that all curves almost overlap on the same parabola, suggesting an identical thermal gradient of the Pt$_1$ electrode through $I_1$ independently of $T_0$, with a small deviation for smaller $T_0$ due to the decrease of $R_\text{Pt}$. In addition, Fig. ~\ref{fig:lsv3}(d) shows $(\underline{\mathscr{T}}_s)^{-1} \equiv ( \mathscr{T}_s / \mathscr{T}_s|_{{I_1} \rightarrow 0} )^{-1}$, the inverse transmission ratio of the spin current generated by the STE normalized by its low current value\footnote{We will consistently use the underlined notation to represent the normalized quantity by the value at the origin. }. The data from the different curves overlap and, similar to the SSE, show a parabolic evolution (see dotted line). This suggests that the primary source of the symmetric nonlinearity between $I_2$ and $I_1$ is simply Joule heating. It therefore justifies the transformation of the current abscissa $I_1$ into a temperature scale $T_1$ in Fig.~\ref{fig:lsv3}(a). Focusing now on the remarkable features of Fig.~\ref{fig:lsv3}(a), one could notice that the low current data taken at $T_0=300$~K fall on a straight line intercepting the origin, as predicted by Eq.~(\ref{eq:dnk}), which is $I_2/I_1 \propto T_1$. Another notable feature, as previously reported\cite{Goennenwein2015,Cornelissen2016}, is that the transmission ratio reaches a maximum at high temperature. 

To support this picture with experimental data, we have plotted in the inset of Fig.~\ref{fig:lsv3} the behavior of $M_T (M_0 - M_T)$ suggested by Eq.~(\ref{eq:dnt}). This should represent the magnon transmission ratio by the high-energy fraction, i.e. the number of available high-energy magnons multiplied by the amount of spin polarization available in the film. We find that the observed variation of $\mathscr{T}_s$ with $T_1$ follows the expected behavior derived from the single temperature variation of the total magnetization shown in the inset Fig.~\ref{fig:lsv3}(b). This provides experimental evidence that the short range behavior is dominated by high-energy magnons and that the density change follows the analytical expression in Eq.~(\ref{eq:dnt}). Furthermore, it is confirmed that the drop in the magnon transmission ratio above 440~K is associated with a drop in the saturation magnetization as one approaches $T_c$, precisely where high-energy magnons reach their maximum occupancy. The drop suggests that high-energy magnons actually prevent STE spin transport. This is the nonlinear deviation expected for a diffusive gas: the higher the number of particles, the more the transport is inhibited (see also Ref.~\cite{kohno_SD}). 
What it shows here is that the magnon transconductance is dominated by high-energy magnons around the emitter. This confirms the initial finding of Cornelissen \textit{et al.}\cite{Cornelissen2015} who drew this conclusion based on the similarity of the characteristic decay of SSE and STE as a function of $d$.

\subsection{Double decay of the magnon transmission ratio} \label{sec:decay}

\subsubsection{Thin films with anisotropic demagnetizing effect}

\begin{figure}
    \includegraphics[width=0.49\textwidth]{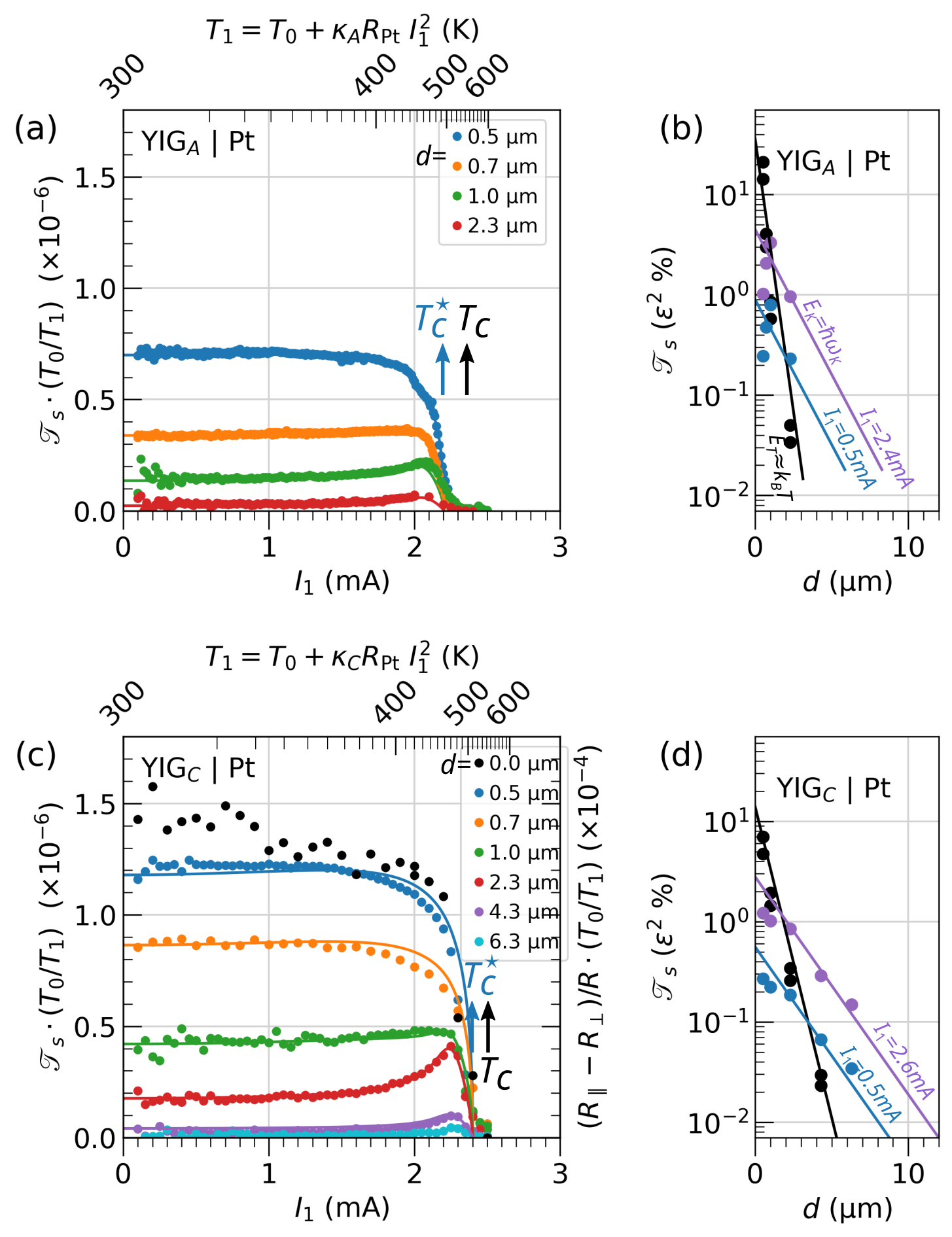}
    \caption{Double exponential spatial decay of the magnon transmission ratio. (a,c) Current dependence of the magnon transmission ratio for (a) the 19~nm thick YIG$_A$ and (c) the 56~nm thick YIG$_C$ thin films. The solid lines are a fit by Eq.~(\ref{eq:2fluid}), where the only variable parameter is the value of $\left . \Sigma_{K}/(\Sigma_{K}+\Sigma_{T}) \right |_{d}$. For the YIG$_C$ sample, we have added in panel (c) the variation of the spin magnetoresistance (right axis), which corresponds to the conductivity at $d=0$. Spatial decay of the magnon transmission ratio for (b) YIG$_A$ and (d) YIG$_C$, respectively. In both cases, the decay of high-energy magnons follows an exponential decay with characteristic length $\lambda_T \approx 0.5\pm0.1$~$\mu$m. The decay of low-energy magnons, on the other hand, follows an exponential decay with characteristic length $\lambda_K=1.5$~$\mu$m for the thinner film (b) and an exponential decay with characteristic length $\lambda_K=1.9$~$\mu$m for the thicker film (d).}
    \label{fig:gap}
\end{figure}

Having established that the spin current is carried by the two-fluids and that the fit allows to extract the respective contributions of high and low-energy magnons, we took a series of experimental data of $\mathscr{T}_s \cdot T_0/T_1$ with different separations $d$ ranging from $0.5 \mu$m to $6.3 \mu$m. The results are shown in Fig.~\ref{fig:gap}. We see directly in Fig.~\ref{fig:gap} that the decay length of the magnon transmission ratio at small $I_1$ is much shorter than the decay length of the magnon transmission ratio at large $I_1$ (spin diode regime). This shows experimentally that each of the two-fluids has a different decay length with $\lambda_T \ll \lambda_K$. These are adjusted by varying $\left . \Sigma_K/(\Sigma_K+\Sigma_T) \right |_d$ while keeping the other parameters in Eq.~(\ref{eq:dnk}-\ref{eq:dnt}). The fits are shown as the solid line in Fig.~\ref{fig:gap}(a,c). The fit parameters are set according to the values given in Table~\ref{tab:fit}. 

By means of the analysis, we obtained the amplitude and the fraction of high-energy vs. low-energy magnons as a function of $d$, which are summarized in panel (b) and extract the two decay lengths $\lambda_K=1.5 \ \mu$m and $\lambda_T=0.4 \ \mu$m, respectively. This confirms the short-range nature of the high-energy magnons and the much longer range of the low-energy magnons.  We note that since the shortest decay length is of the same order of magnitude as the spatial resolution of standard nanolithography techniques, the regime of magnon conservation could probably never be achieved in lateral devices. 
Note that there is the discrepancy that the vanishing of $I_2$ occurs slightly before $T_c$. We will show that this occurs systematically on all our samples (see subsection $\textit{3}$). 
The same analysis applied to the YIG film with larger thickness (panels (c,d)) reveals an identical behavior of the high-energy magnons, whereas the decay of the low-energy magnons is slightly slower with $\lambda_K = 1.9\ \mu$m.

\begin{table*}[t]
  \caption{Fitting parameters by Eq.~(\ref{eq:2fluid}).}
  \begin{ruledtabular}
    \begin{tabular}{*{9}{c}}
     & $t_\text{YIG}~(\rm nm)$ & $n_\text{sat}$ & $T_c^\star$~(K) & $\mathcal{I}_{\text{th},0}$~(mA)  & {$\lambda_K$  ($\mu$m)} & {$\lambda_T$  ($\mu$m)}  & {$\Sigma_{K,0}^{L\rightarrow R}$}  &  {$\Sigma_{T,0}^{L\rightarrow R}$} \\ \hline
      
      YIG$_A$ & 19 & 4 & 495 & 8 &
      {1.5} & 
      {0.4}  & {5 $\%$} & {37 $\%$} \\

      (Bi-)YIG$_B$ & 25 &  11 & 480 & 3 &
      {3.8} & 
      {0.5} & {4 $\%$} & {39 $\%$} \\
      
      YIG$_C$ & 56 &  4 & 515 & 8 &
      {1.9} & 
      {0.6}  & {3 $\%$} & {15 $\%$} \\
      
      YIG$_D$ & 65 &  4 & 545 & 8 &
      {} & 
      {} & {}  & {} \\

\end{tabular}
\end{ruledtabular}\label{tab:fit}
\end{table*}

We do not see an obvious increase in the transmission ratio in thinner films (YIG$_A$), although Eq.~(5) of Ref.~\cite{kohno_SD} predicts inverse proportionality as previously observed experimentally \cite{Shan2016}, which can be attributed to the difference in material quality. Nevertheless, an interesting feature observed when comparing Fig.~\ref{fig:gap}(a) and (c) is that the ratio of low-energy magnons to high-energy magnons increases with decreasing film thickness. This can be attributed to an increase in the cutoff wavevector, where the magnons behave two-dimensionally, and thus the spectral range, where the density of state remains constant, which favors the exposure of the increasing occupancy of low-energy magnons. The longer decay length in the thicker film is also consistent with the longer propagation distance expected for ballistic low-energy magnons, whose propagation range is determined by the film thickness. However, the enhancement is not proportional to the thickness, suggesting that some other undefined process is also involved in this decay.

We emphasize that the shape of the decay observed in Fig.~\ref{fig:gap}(b) and (d) corresponds to a double exponential decay with two different decay lengths in unprocessed data. This reinterprets the double decay behavior reported in previous nonlocal transport measurements\cite{Cornelissen2015,cornelissen2016magnetic,Shan2016,gomez2020differences}. The interpretation presented in this work is different from the one proposed by Cornelissen \textit{et al.}, where it was related to the boundary condition of the diffusion problem\cite{Cornelissen2015}. We note that while changing the current bias $I_1$ can affect the ratio between the two-fluids, it does not change the decay length, as shown by the purple lines in panel (b,d). This is consistent with the notion that the bias affects the mode occupation of the transported magnons but not their character. The obtained decay lengths are in rough agreement with the expected decay length of these two populations as discussed in Sect.\ref{sec:analytical}. Note also that the high- and low-energy magnon length scales appear to be similar to the energy and spin relaxation length scales observed in the Spin Seebeck effect as proposed by A. Prakash et al. \cite{Prakash2018}, the correlation between the length scales is a complex issue that warrants a more rigorous theoretical investigation (see also conclusion below).

We note that our value of $\lambda_K$ appears to be dependent on thickness and anisotropy (see Table~\ref{tab:fit}). This contradicts the behavior observed for thicker films ($t_{\mathrm{YIG}}>200$~nm), where the value was reported to be independent of film thickness\cite{Shan2016}. The latter observation may be consistent with the assignment of the dominant low-energy propagating magnons to the $E_{\overline{K}}$ mode (orange dot in Fig.~\ref{fig:vg}). We believe that the group velocity there is weakly dependent on $t_\text{YIG}$, at least for thick films (see discussion above).  We should emphasize here that our report does not cover the same dynamic range as those reported in thicker films, due to the lower signal-to-noise ratio. It is possible that a third exponential decay could appear at much lower signal levels. A possible explanation for the long range behavior could be that the angular momentum is carried by circularly polarized phonons, which have been found to have very long characteristic decay lengths in the GHz range\cite{An2020,an2022bright}.

Finally, it is useful to quantify the spin current emitted by the STE, as shown in panel (b,d). Renormalizing the transmission ratio coefficient $\mathscr{T}_s$ by the product of the spin transfer efficiency at both the emitter and collector interfaces, $\epsilon_1 \cdot \epsilon_2$ (see Table~1 of Ref.~\cite{kohno_SD}), we observe that only 10\% of the generated magnons reach a collector placed at $d=0.2\, \mu$m away. This percentage increases to 15\% by extrapolating the decay to $d=0$, which is the proportion of \emph{itinerant} magnons among the total generated, and there are about an order of magnitude ($\times 14$) more high-energy magnons than low-energy magnons below the emitter. Taking into account the fact that magnons can escape from both sides of the emitter, while we monitor only one side, we can estimate that 70\% of the generated magnons remain localized. This \emph{localization} is the consequence of three combined effects, which mainly affect the low-energy magnons: \textit{i)} STE primarily favors an increase in density at the bottom of the magnon manifold, which has zero group velocity \textit{ii)} STE, as an interfacial process, efficiently couples to surface magnetostatic modes \cite{eshbach1960surface}, The nonlinear frequency shift associated with the demagnetizing field \cite{divinskiy2019controlled} produces a band mismatch at high power between the region below the emitter and the outside, which prevents the propagation of magnons (see part I~\cite{kohno_SD}). The spatial localization could be induced either by the thermal profile of the Joule heating \cite{an2021short} or by the self-digging ball modes \cite{Demidov2012,ulrichs2020chaotic,schneider2021stabilization}. This rational concerns mainly the magnons whose wavelengths are shorter than the width of the Pt electrode.

Another confirmation is the variation of the ratio between low-energy magnons and high-energy magnons with the uniaxial anisotropy. When the latter compensates the out-of-plane depolarization field, we observe a suppression of the low-energy magnon confinement, and the transmitted signal at large distances (10~$\mu$m) fully replicates the variation of low-energy magnons under the emitter.

\subsubsection{Thin films with isotropically compensated demagnetizing effect}

\begin{figure}
    \includegraphics[width=0.49\textwidth]{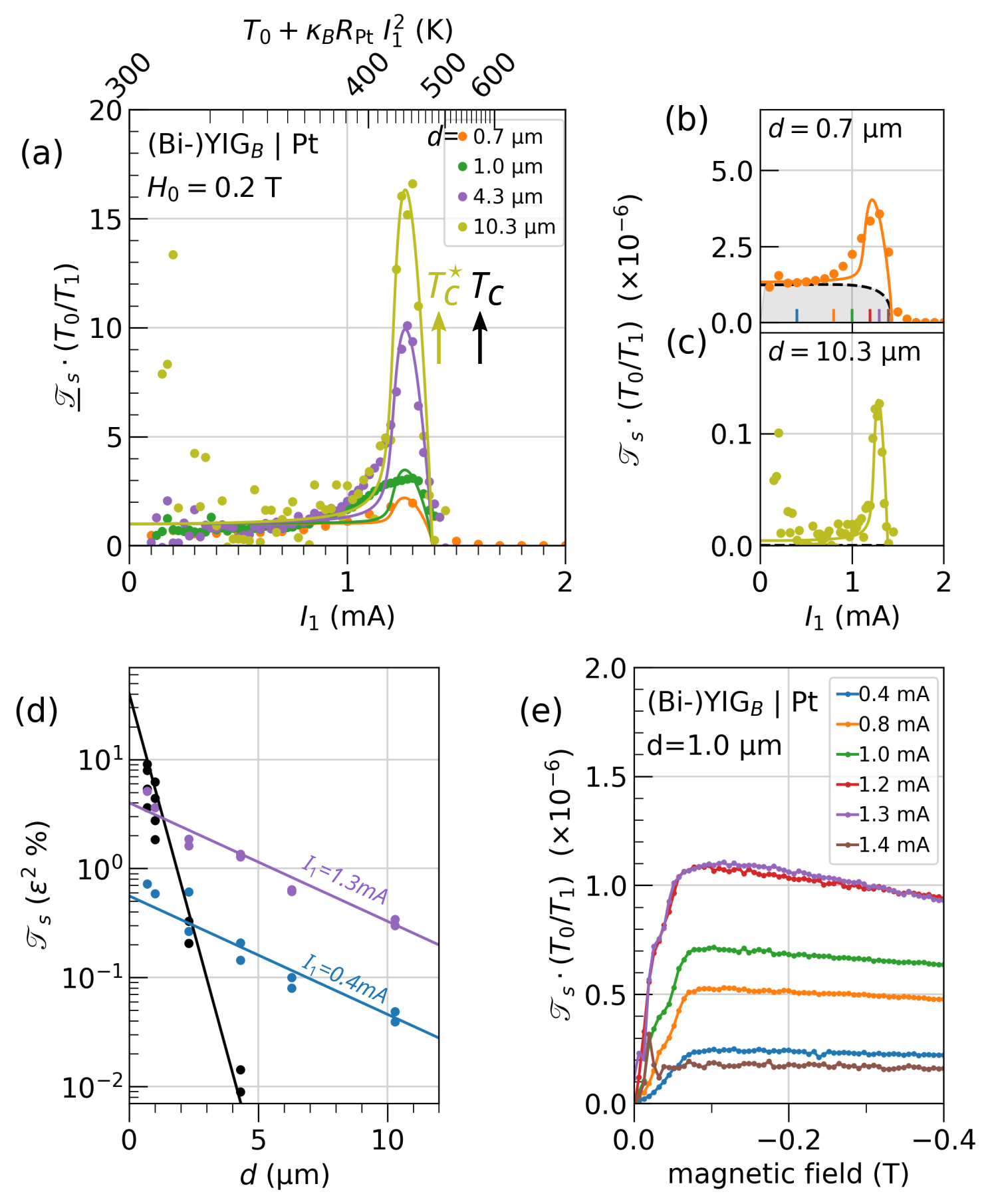}
    \caption{Two-fluid behavior in thin films with isotropically compensated demagnetization effect ($M_\text{eff}=0$). (a) Variation of the spin diode signal $\mathscr{T}_s$ measured in BiYIG$_B$ for different emitter-collector separations $d$. The main panel (a) shows the normalized magnon transmission ratio as a function of $T_1$, while the right panels show the corresponding current dependence for (b) $d=0.7~\rm \mu m$ and (c) $d=10.3~\rm \mu m$. The solid lines are fits by Eq.~(\ref{eq:2fluid}), with the only variable parameters, $\Sigma_K$ and $\Sigma_T$, representing the fraction of low and high-energy magnons.  (d) Spatial decay of the two-fluid model separating the contributions of high-energy and low-energy magnons. The observed decay can be explained by a short decay $\lambda_T \approx 0.5$~$\mu$m of the high-energy magnon contribution ($k_B T$, black line) and a long decay $\lambda_K \approx 4.0$~$\mu$m of the low-energy magnon contribution ($\hbar \omega_K$, magenta and blue lines). The data at $I_1=1.3~\rm mA$ show the decay behavior in the condensed regime. (e) Magnetic field dependence of the normalized magnon transmission ratio at different currents.}
    \label{fig:lsv5}
\end{figure}

In this section, we will clarify the influence of self-localization on the saturation threshold $n_\text{sat}$ that we introduce in our analytical model. For this purpose, we have repeated the experiment on a Bi-YIG$_B$ sample. This material has a uniaxial anisotropy corresponding to the saturation magnetization (see Table~1 in Ref.~\cite{kohno_SD}). As a consequence, the Kittel frequency follows the paramagnetic proportionality relation $\omega_K = \gamma H_0$ (similar to the response of a sphere), where the value of $\omega_K$ is independent of $M_T$ and the cone angle of precession, and therefore exhibits a vanishing nonlinear frequency shift\cite{divinskiy2019controlled,Evelt2018,guckelhorn2021magnon} (see further discussion in Ref.~\cite{kohno_SD}). We refer to this as an isotropically compensated material. We emphasize, however, that although the nonlinear frequency shift is zero, the system is still subject to saturation effects\cite{gurevich2020magnetization}. Compensation of the out-of-plane demagnetization factor eliminates only the ellipticity of the trajectory caused by the finite thickness, but not the self-depolarization effect of the magnons on themselves. The latter depends on the angle between the propagation direction and the equilibrium magnetization direction and is the origin of the magnon manifold broadening. 

As shown in Fig.~\ref{fig:lsv5}(a), the nonlinear behavior of $\mathscr{T}_s$ observed in the Bi-YIG$_B$ sample is qualitatively similar to that of YIG$_C$. Quantitatively, however, the magnitude of the spin diode effect is more pronounced in the former case. This is especially noticeable at long distances. Comparing Fig.~\ref{fig:lsv5}(b) ($d=0.70$~$\mu$m) with Fig.~\ref{fig:lsv5}(c) ($d=10.3$~$\mu$m), for the former the conductivity can only be increased by a factor of 3 with respect to its initial value, while for the latter it can be increased by a factor of $~15$. This is again due to the filtering out of the background of high-energy magnons: in the case of large distances, the contribution of low-energy magnons is more pronounced. Recalling that in YIG$_C$ the conductivity was enhanced by a factor of 7 by low-energy magnons (see $d>4.3$~$\mu$m data in Fig.~\ref{fig:gap}(c) or Fig.~~7 of Ref.~\cite{kohno_SD}), here a larger fitting parameter of $n_\text{sat}=11$ is used in Bi-YIG$_B$ while $n_\text{sat}=4$ is used in YIG$_C$, indicating a larger threshold for saturation. This is consistent with the suppression of the nonlinear frequency shift affecting the long wavelength spin wave in the YIG$_C$ sample. This result suggests that removing the self nonlinearity on the long wavelength magnons improves the ability to generate more propagating magnons. It can also be understood as the removal of the self-digging process under the emitter in pure YIG samples. The fit parameters are listed in Table~\ref{tab:fit}. Note that the discrepancy between $T_c$ and $T_c^\star$, which marks the drop of $\mathscr{T}_s$, is even more pronounced in this system. The drop occurs 70~K below $T_c$. We will return to this point in the last subsection.
 
In Fig.~\ref{fig:lsv5}(d) we plot the spatial decay of $\mathscr{T}_s$ renormalized by $\epsilon^2$, obtained from fits with Eq. ~(\ref{eq:2fluid}) in percent for high-energy magnons in black, low-energy magnons at $I_1=0.4$~mA ($\mu_m \ll E_g$) in blue, and $I_1=1.3$~mA ($\mu_m \approx E_g$) in purple. The two decay lengths are $\lambda_T \approx 0.4 \mu$m for high-energy magnons, in agreement with the results in YIG, and a much larger value of $\lambda_K = 4$~$\mu$m for low-energy magnons. The latter value is similar to the decay length of low-energy magnons observed by BLS in these films\cite{Evelt2018}. Moreover, it is in good agreement with the estimate made in Sect.~\ref{sec:analytical}.

For the sake of completeness, we plot the magnetic field dependence for different $I_1$ in Fig.~\ref{fig:lsv5}(e). The decrease of the signal at zero field is due to the residual out-of-plane anisotropy, which forces the magnetization to be along the film normal, resulting in no STE applied by Pt. The magnon transmission ratio becomes maximum near 0.05~T, which is the saturation of the effective magnetization for BiYIG. The field dependence at a larger field than 0.05~T becomes significant for the current values near the appearance of the peak in (a) at $I_1=1.3$ mA, where the conductivity of low-energy magnons reaches the highest. As noted in a previous study\cite{thiery2018}, the fact that we see a dependence with magnetic fields is direct evidence that we are dealing here with low-energy magnons. Here the extra sensitivity of $\mathscr{T}_s$ to changes in $\mathcal{I}_\text{th}$ near $I_\text{pk}$, as discussed above in the context of describing the behavior of Fig.~\ref{fig:hdep}, is clearly illustrated here with the BiYIG sample.  

\begin{figure}
    \includegraphics[width=0.49\textwidth]{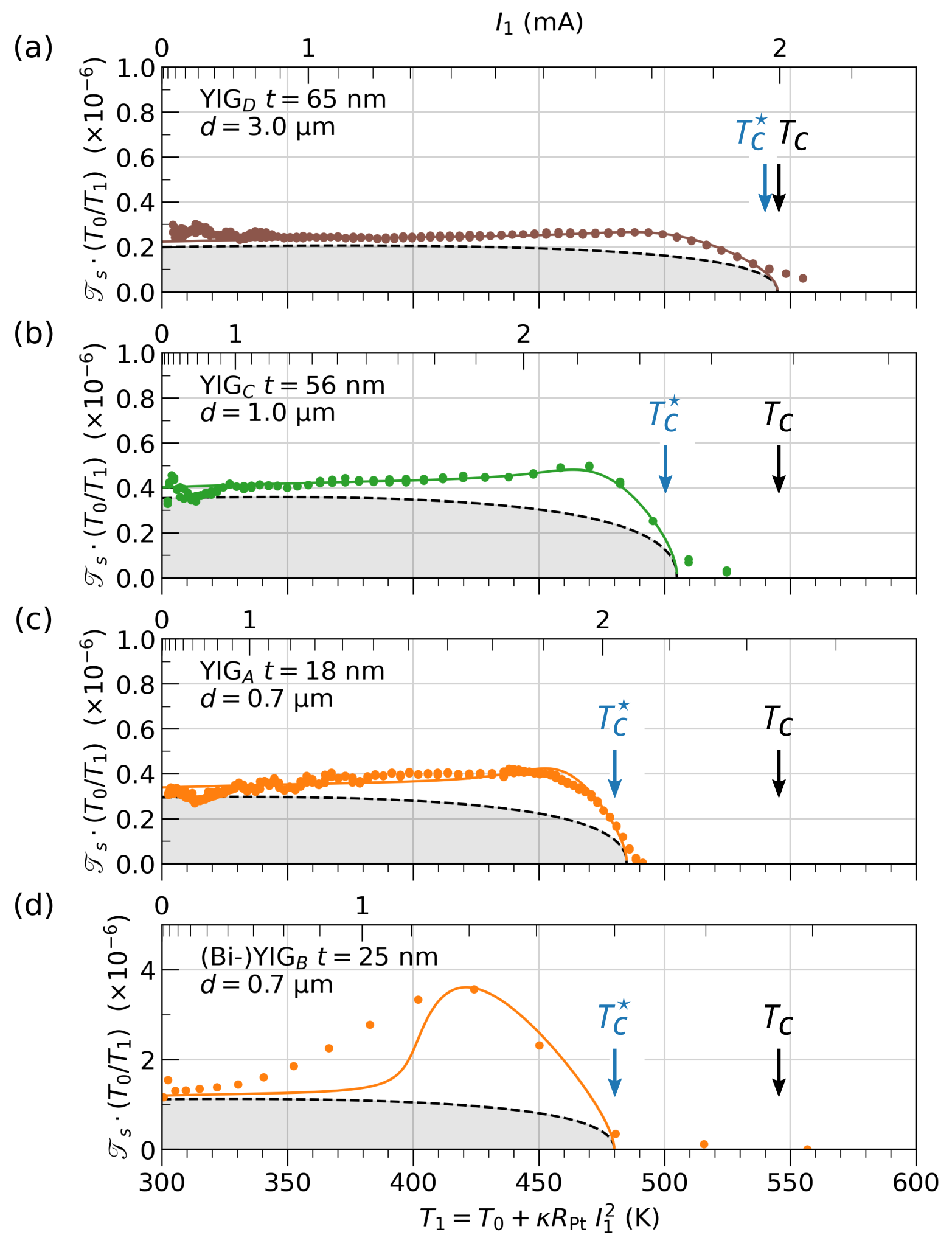}
    \caption{Dependence of $T_c^\star$ on the thickness of YIG films. Comparison of nonlocal devices with approximately the same ratio of high-energy magnons to low-energy magnons at $I_1 \rightarrow 0$. We observe an increase in $T_c -T_c^\star$ with decreasing film thickness, suggesting an increasing influence of low-energy magnons at high power $I_1 \rightarrow I_c$ with decreasing film thickness.}
    \label{fig:tc}
\end{figure}

\subsubsection{Discrepancy between $T_c$ and $T_c^\star$}

Finally, we discuss the disappearance of the magnon transmission ratio already at $T_c^\star$ far below the experimentally determined $T_c$ (see Fig.~\ref{fig:carac}). We note in Fig.~\ref{fig:gap} and Fig.~\ref{fig:lsv5} that all curves collapse at the same value independent of $d$. This clearly points to a problem that only concerns the region below the emitter, since there is a lateral temperature gradient. To this end, we summarize the normalized magnon transmission ratio for YIG samples as a function of emitter temperature $T_1$ in Fig.~\ref{fig:tc} with different thicknesses. To avoid any influence of thermal gradients, we have chosen devices whose spacing $d$ leads to a similar ratio between $\Delta n_T$ and $\Delta n_K$. This requires $d$ to increase with increasing film thickness, suggesting a decreasing contribution of low-energy magnons. We speculate that the collapse can be caused either by the onset of strong electron-magnon scattering as the YIG film becomes conducting\cite{Thiery2018a,schlitz2021nonlocal}, or by a reversal of the equilibrium magnetization below the emitter, which becomes aligned with the injected spin direction\cite{avci2017current,ulrichs2020chaotic}. In the latter case, the magnetization below the emitter and collector are opposite, suppressing any spin transport. This process is consistent with the assumption that a large fraction of the injected spins remain localized. This process is also consistent with the decrease of $\overline{\mathscr{V}_2}$ observed at large $I_1$, where now the electric current decreases the effective temperature of the spin system (decrease fluctuations) despite the fact that $I_1 \cdot H_x < 0$.

We examine the other clues that support this picture. If one compares the discrepancy between $T_c^\star$ and $T_c$ between the different samples, one can clearly see on the data in Fig.~\ref{fig:tc} that the discrepancy increases with decreasing film thickness, as expected for an increased surface effect of STE and reduced volume of polarized spins. Another indication is the fact that the largest discrepancy is observed on films with large uniaxial anisotropy, as shown in Fig.~\ref{fig:lsv5}(a). This is in agreement with the observation made on nano-devices on the switching of the magnetization direction by the spin Hall effect\cite{Miron11}. Nevertheless, the discrepancy does not seem to scale simply with $t_\text{YIG}$ in our observation, suggesting that there may be additional phenomena at play that are responsible for the vanishing magnon transmission ratio at high temperature while the system is still in its ferromagnetic phase (see also the discussion of Fig.~5 of Ref.~\cite{kohno_SD}).

We have tentatively calculated $I_f$, the critical current required to flip the magnetization. We call $n_\text{sat} = V M_1 / (\gamma \hbar)$ the total number of spins that remain polarized under the emitter. We compare this to the number of injected spins within the spin-lattice relaxation time, which is $I_f \epsilon / (2 e \alpha_\text{LLG} \omega_K)$. Equalizing the two quantities, we find that $I_f = 2.5$~mA for YIG$_A$ samples. According to the upper scale of Fig.~\ref{fig:gap}, $T_c$ is reached when $I=2.7$~mA. Using Fig.~\ref{fig:carac}, we can calculate the temperature difference produced by Joule heating between these two values, and the result is about 65~K. This is very close to the shift of 50~K observed experimentally on this sample. 

While there are indications that a shift occurs, and the number roughly matches the expected numbers, the above paragraph is still rather speculative at this stage, and a direct proof is still missing.  For the sake of completeness, it is worth mentioning that there may be alternative explanations. One possibility is a decrease of $T_c$ in the region below the Pt. The origin of such an effect could be interdiffusion of Pt atoms inside the YIG at the interface. More thorough systematic studies will be required to clarify this point.

\section{Conclusion}

Through these two consecutive reviews, we present a comprehensive picture of magnon transport in extended magnetic insulating films, covering a wide range of current and magnetic field bias, substrate temperature, as well as nonlocal geometries with varying propagation distance. The picture of the two-fluid model expressed in this part II, complemented by a picture of the nonlinear behavior of the low-energy magnon expressed in part I, is formulated analytically and it is supported by a series of different experiments that include nonlocal transport on different thicknesses YIG thin films with different garnet composition, different interfacial efficiency, as well as different thermalization. While providing a comprehensive study of these materials, our model accounts for almost all the experimental observations within this common framework. 

What the analytical model allows to do is:
\begin{enumerate}[label=\textit{\roman*})]
    \item to describe the expected signal in the linear regime [Eq.~(6) in part I]
    \item to fit the nonlocal transport data well on the whole current range and for different separation between the electrodes using very few parameters ($\mathcal{I}_\text{th,0}$, $n_\text{sat}$, $T_c^\star$, $\lambda_T$, $\lambda_K$, $\Sigma_T$ and $\Sigma_K$)
    \item to incorporate all relevant physical effects: effect of Joule heating on $M_1$, divergent form of magnon-magnon relaxation.
\end{enumerate}

What it doesn't do, but could be important:
\begin{enumerate}[label=\textit{\roman*})]
    \item to take into account the propagation properties (propagation angle, group velocity, mode selection by the electrode geometry, spatial variation of these properties due to the temperature gradient) of the magnons excited under the emitter to know how they contribute to the signal under the collector. 
    \item to take into account nonlinear magnon localization effects under the emitter (for YIG in particular).
    \item to take into account the effects of high power (change in temperature or change in low energy magnon occupancy) on damping, exchange constant (and thus group velocity), pumping, and detection efficiency.
\end{enumerate}

The fact that these points are not directly considered and that the fits are excellent means that these effects are effectively used in the other components of the model. In particular, Eq. (6) of the relaxation in part I is very general and can absorb many different physical effects, hence the effectiveness of the model.

In this paper, we assume that low-energy magnons propagating in the ballistic regime lead to a magnon transconductance that follows an exponential spatial decay in thin film geometries. This argument follows from the experimental finding that in all BLS experiments monitoring the low-energy part of the magnon manifold, the amplitude of the signal follows an exponential decay. Nevertheless, the transport behavior in the clean limit, where the magnon mean free path is larger than the sample boundary, is in itself a very interesting line of research. 

Another open question concerns the premature collapse of the signal at $T_c^\star$. We have tentatively explained this as a potential switching of the magnetization direction below the emitter. However, direct evidence for such a process remains elusive. We think that spin transport in materials with low magnetization or close to the paramagnetic phase are both very interesting topics.

Finally, we summarize the main result of our two-fluid model, which separates the low-energy magnons from the high-energy ones. This allows us to propose an alternative explanation for the measured variation of the magnon transmission ratio with distance, due to a double exponential decay. Each of the fluids has its own transport characteristics, which are expressed by two different propagation lengths. A decay length in the submicron range is assigned to the high-energy magnon and a decay length above the micron range is assigned to the low-energy magnon. This explanation implies that even in the short-range regime, the magnon number is not a conserved quantity, and thus any analogy to electronic transport should take this rapid decay into account. Despite the fact that the model includes several parameters, there are still open questions. The similarity of the decay of SSE and STE currents with $d$ must be reconciled with our results. A possible reason is that low-energy magnons participate in the SSE transport in the long range\cite{kikkawa2015critical}. Although the amount of quanta carried is clearly $E_T/E_K \sim 10^3$ against the latter, we should keep in mind that we are dealing with a tiny signal. The role of acoustic phonons \cite{An2020,an2022bright} in this process is still unclear. Recent experiments have shown that they are strongly coupled to low-energy magnons and also benefit from a very low decay length. Of particular interest is the contribution of circularly polarized acoustic phonons, which have been shown to be strongly coupled to long-wavelength spin waves while allowing angular momentum transfer over large distances.

\begin{acknowledgments}
This work was partially supported by the French Grants ANR-18-CE24-0021 Maestro and ANR-21-CE24-0031 Harmony;  the EU-project H2020-2020-FETOPEN k-NET-899646; the EU-project HORIZON-EIC-2021-PATHFINDEROPEN PALANTIRI-101046630. K.A. acknowledges support from the National Research Foundation of Korea (NRF) grant (No. 2021R1C1C201226911) funded by the Korean government (MSIT). This work was also supported in part by the Deutsche ForschungsGemeinschaft (Project number 416727653).

\end{acknowledgments}


\newpage

\section{Annex}

\setcounter{figure}{0}
\renewcommand{\thefigure}{S\arabic{figure}}%

\begin{figure}
    \includegraphics[width=0.49\textwidth]{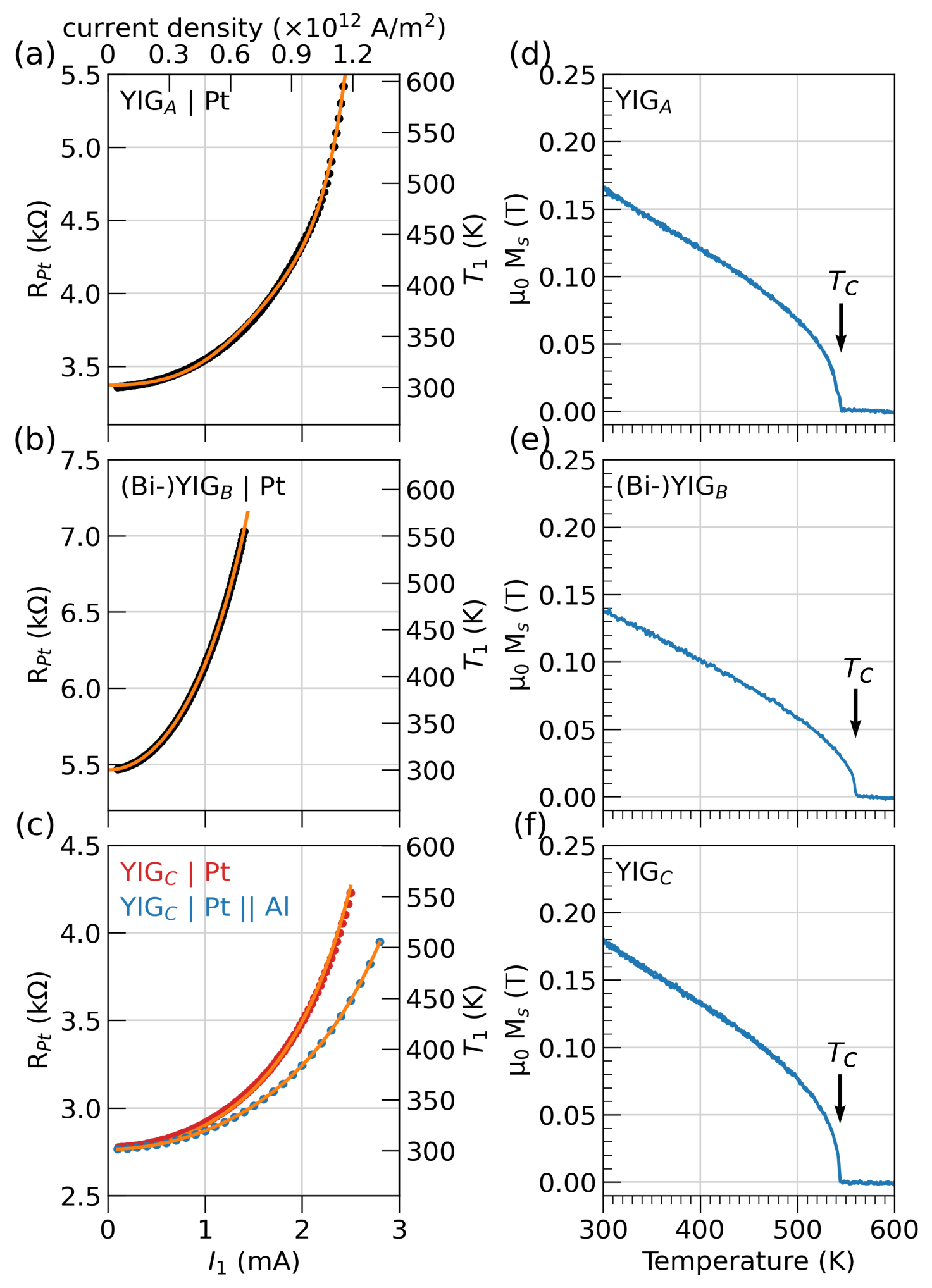}
    \caption{Characterization of garnet thin films. The left column (a,b,c) shows the variation of the Pt resistance as a function of the injected current for YIG$_A$, (Bi-)YIG$_B$ and YIG$_C$ without and with Al coating, respectively (see Table~1 of Ref.~\cite{kohno_SD}). The right ordinate allows to convert the current bias into a temperature increase in the range [300,600]~K due to Joule heating. The upper abscissa gives the corresponding current density in Pt. The right column (d,e,f) shows the corresponding variation of the saturation magnetization in the [300,600]~K range.}
    \label{fig:carac}
\end{figure}

\subsection{Sample characterisation}

The 4 magnetic garnet films (see Table~\ref{tab:fit}) used in this study have been grown by 2 different methods: liquid phase epitaxy in the case of YIG$_{A,C,D}$ and pulsed laser deposition in the case of (Bi-)YIG$_B$. Their macroscopic magnetic properties have been characterized using a commercial vibrating sample magnetometer, where the sample temperature can be controlled by a flow of argon gas from room temperature to 1200K. Curves of magnetization versus temperature in the range of 300K to 600K are shown in Fig.~\ref{fig:carac}(d-f). They highlight the value of the Curie temperature ($T_c$) for each sample summarized in Table~1 in Ref.~\cite{kohno_SD}. Similarly, the Pt metal for the middle electrode was deposited by 2 different techniques: e-beam evaporation in the case of YIG$_A$ and YIG$_C$ and sputtering in the case of (Bi-)YIG$_B$.

In this work we convert the Joule heating associated with the circulation of an electric current $I_1$ in the emitter into a temperature increase, which we plot on the abscissa of Fig.~\ref{fig:lsv3}, Fig.~\ref{fig:tc} and Fig.~6, Fig.~7 of Ref.~\cite{kohno_SD}. This is done by calibrating $\left . R_\text{Pt} \right |_{I_1}$: the variation of the resistance Pt$_1$ with the injected electric current $I_1$. We introduce the calibration factor
\begin{equation}
\kappa_{A,B \text{ or } C} =\kappa_\text{Pt} \dfrac {R_\text{Pt}/R_0 -1}{R_\text{Pt} I_1^2} \hspace{0.2cm}, \label{eq:kappa}
\end{equation}
for the conversion coefficient, with $R_0 \equiv \left . R_\text{Pt} \right |_{I_1=0} = \rho_\text{Pt} L_\text{Pt}/(w_1 t_\text{Pt})$ is the nominal value of the Pt wire resistance and the coefficient $\kappa_\text{Pt} = R_\text{Pt}/\partial_T R_\text{Pt}$ is obtained by monitoring the variation of the Pt resistance at low current vs. substrate temperature. The obtained values of $\kappa_\text{Pt}$ and $\rho_\text{Pt}$ are given in Table~1 in Ref.~\cite{kohno_SD}. Fig.~\ref{fig:carac}(a-c) shows the R-I curves with corresponding temperature considering Joule heating for each sample.

\bibliography{newlib}
\end{document}